\documentclass{WileyMSP-template}
\usepackage{amsmath}
\usepackage{amsfonts}
\usepackage{amssymb}
\usepackage{hyperref}
\usepackage{graphicx}
\usepackage{color}
\usepackage{amsmath}
\usepackage{amssymb}
\usepackage{tabularx}
\usepackage[utf8]{inputenc}
\usepackage[table]{xcolor} 
\usepackage{siunitx} 
\usepackage[utf8]{inputenc}
\usepackage[T1]{fontenc}
\usepackage{soul}

\DeclareRobustCommand{\wh}[1]{\ensuremath{\widehat{#1}}}

\newcommand{\ket}[1]{\left| #1\right\rangle}
\newcommand{\kett}[1]{|#1\rangle\!\rangle}

\newcommand\Tr{\mathrm{Tr}}

\newcommand{\er}{{\bf r}}
\newcommand{\re}{{\rm e}}
\newcommand{\mL}{\widehat{\mathcal{L}}}
\newcommand{\dL}{{\dagger L}}
\newcommand{\dR}{{\dagger R}}

\newcommand{\change}[1]{{#1}}

\begin{document}

\pagestyle{fancy}
\rhead{\includegraphics[width=2.5cm]{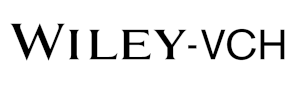}}

\title{\change{Quantization of} Polaritons Confined in Dielectric Structures}

\maketitle

% Author: Please give full first and last names for authors and include * after the name of all corresponding authors

\author{Amir Rahmani}
\author{Dogyun Ko}
\author{Maciej Dems}
\author{Andrzej Opala}
\author{Michał Matuszewski*}

% Dedication

% Affiliations: Please provide adacemic titles (Prof. or Dr.) for all authors where applicable, and include an institutional email address for all corresponding authors
\begin{affiliations}

Amir Rahmani, Dogyun Ko, Andrzej Opala, Michał Matuszewski\\
Institute of Physics Polish Academy of Sciences, Al. Lotnik\'{o}w 32/46, 02-668 Warsaw, Poland\\

Amir Rahmani, Dogyun Ko, Andrzej Opala, Michał Matuszewski\\
	Center for Theoretical Physics, Polish Academy of Sciences, Al. Lotnik\'{o}w 32/46, 02-668 Warsaw, Poland\\
Email Address: mmatuszewski@cft.edu.pl\\
Maciej Dems\\
Institute of Physics, Lodz University of Technology, ul. Wólczańska 217/221, 93-005 Łódź, Poland\\
Andrzej Opala\\
Institute of Experimental Physics, Faculty of Physics, University of Warsaw, ul. Pasteura 5, PL-02-093 Warsaw, Poland\\

\end{affiliations}

% Keywords: Please provide a minimum of three and a maximum of seven keywords, separated by commas


% Abstract should be written in the present tense and impersonal style (i.e., avoid we), and be at most 200 words long
\begin{abstract}

Light-matter interaction in the regime of strong quantum coupling is usually treated within the framework of the Hopfield model. However, the picture of coupling well-defined modes of light and matter is correct only as long as the shapes of these eigenmodes are not substantially modified by the interaction. Moreover,  parameters of theoretical models are usually obtained by fitting to experimental data. To date, there is no straightforward method to determine a quantum master equation corresponding to a system with specific dielectric structure, which may lead to incompatibility of theoretical descriptions and physical realizations. In this work, a recipe for obtaining a quantum model in the polariton eigenmode basis  is presented, based on Bogoliubov transformation in the conservative case and third quantization technique in the dissipative case. It is shown how this method can be used for boosting interaction strength and engineering nonlocal many-body interactions  in carefully designed nanostructures, resulting in strongly nonclassical correlations of emitted light.

\end{abstract}

\section{Introduction} 

\change{The idea of strong light-matter interaction dates back to early 1950s, when first theoretical concepts were formulated~\cite{tolpygo1950physical,huang1951lattice,Hopfield_Polaritons,Pekar1957,Agranovich1957}}. In the seminal paper \change{published in} 1958, John~J.~Hopfield proposed a model of photons interacting with excitons in the regime of strong coupling, where the system is quantized in terms of hybrid light-matter particles named polaritons~\cite{Hopfield_Polaritons}. Since then, experiments have confirmed this theory in diverse settings. The polariton family~\cite{basov2020polariton} now includes, among others, photons strongly coupled to excitons~\cite{weisbuch1992observation,orfanakis2022rydberg,Kang2023}, phonons~\change{\cite{kowalski2025ultraconfined,passler2018strong,vicentini2026real,galiffi2024extreme,Luo2024}}, plasmons~\cite{Pres2023}, polarons~\cite{ravets2018polaron}, magnons~\cite{Zhang2017}, Rydberg~\cite{Kim2021,jia2018strongly} and ultracold atoms~\cite{kwon2022formation}, existing in various materials and geometries including cavities~\cite{kavokin2017microcavities}, waveguides~\cite{Kędziora2024,liran2024electrically}, nanorods~\cite{VanVugt_Nanowire}, nanotubes~\cite{Graf2016}, two-dimensional materials~\cite{zhang2021van}, moir\'e structures~\cite{zhang2021van}, and living systems~\cite{coles2017polaritons}. Depending on the physical platform, they exhibit unique physical properties such as extremely low effective mass~\cite{RevModPhys.82.1489}, bosonic condensation~\cite{Snoke_BEC},  hyperbolic dispersion~\change{\cite{kowalski2025ultraconfined,galiffi2024extreme,ruta2023hyperbolic}}, ultrastrong coupling~\cite{frisk2019ultrastrong}, or strong interactions~\cite{estrecho2019direct}. Applications of polaritons include ultralow threshold lasing~\cite{schneider2013electrically,christopoulos2007room}, topological lasers~\cite{Klembt2018,Comaron_Topological}, optical logic~\cite{Opala_23},  neurmorphic computing~\cite{Opala_19,Ballarini2020,Mirek_2021,matuszewski2021energy,Opala_23,kavokin2022polariton} and quantum simulations~\cite{boulier2020microcavity}.

Polariton systems are usually well described by the Hopfield Hamiltonian, which in the rotating wave approximation (RWA) reduces to the simple eigenmode problem
\begin{align}
     \label{eq:Hopfield}\begin{pmatrix}
         E_C(\mathbf{k}) & \Omega/2 \\ \Omega/2 & E_X 
     \end{pmatrix}\begin{pmatrix}
         C \\
         X
     \end{pmatrix}
     = E(\mathbf{k}) \begin{pmatrix}
         C \\
         X
     \end{pmatrix},   
\end{align}
where $C,X$ are quantum amplitudes of photons and excitons (or other matter modes) in the polariton eigenstate, $\mathbf{k}$ is the momentum, $E_C(\mathbf{k})$ is the photon energy dispersion, $E_X$ is  matter excitation energy (here constant), and $\Omega$ is the Rabi energy quantifying the light-matter interaction. The resulting upper and lower polariton mode energies are 
\begin{align} \label{eq:Epm}E_\pm(\mathbf{k})&=\frac{E_C(\mathbf{k})+E_X}{2}\pm\frac{1}{2}\sqrt{(E_C(\mathbf{k})-E_X)^2+\Omega^2}.
\end{align}
This model is also valid in systems that are partly confined, such as microcavities \change{and quantum wells~\cite{andreani1991radiative,savona1994quantum,savona1995quantum,ivchenko1991excitonic}}, where the momentum $\mathbf{k}$ is replaced by the momentum parallel to the cavity plane $\mathbf{k}_\parallel$. While the predictions of this model are often very accurate, the parameters are almost always obtained by fitting to the experimental data. On the other hand, one can ask when  this simple model is a correct description of physical systems. The most obvious case which goes beyond its applicability is when more than two modes of light or matter interact, but in this case a simple multimode extension of~\eqref{eq:Hopfield} usually suffices. When translational variance is absent, $\mathbf{k}$ is not a good quantum number, but as long as spatial modes can be well defined using another quantum number, Equation~\eqref{eq:Hopfield} can be still used.  It is also possible to include polarization degrees of freedom~\cite{shelykh2009polariton}, effects of dissipation and pumping~\cite{wouters2007excitations} and ultrastrong light-matter coupling (beyond RWA)~\cite{Hopfield_Polaritons,todorov2010ultrastrong} by suitable extensions of the Hopfield model. 

\begin{figure}
    \centering
    \includegraphics[width=0.5\linewidth]{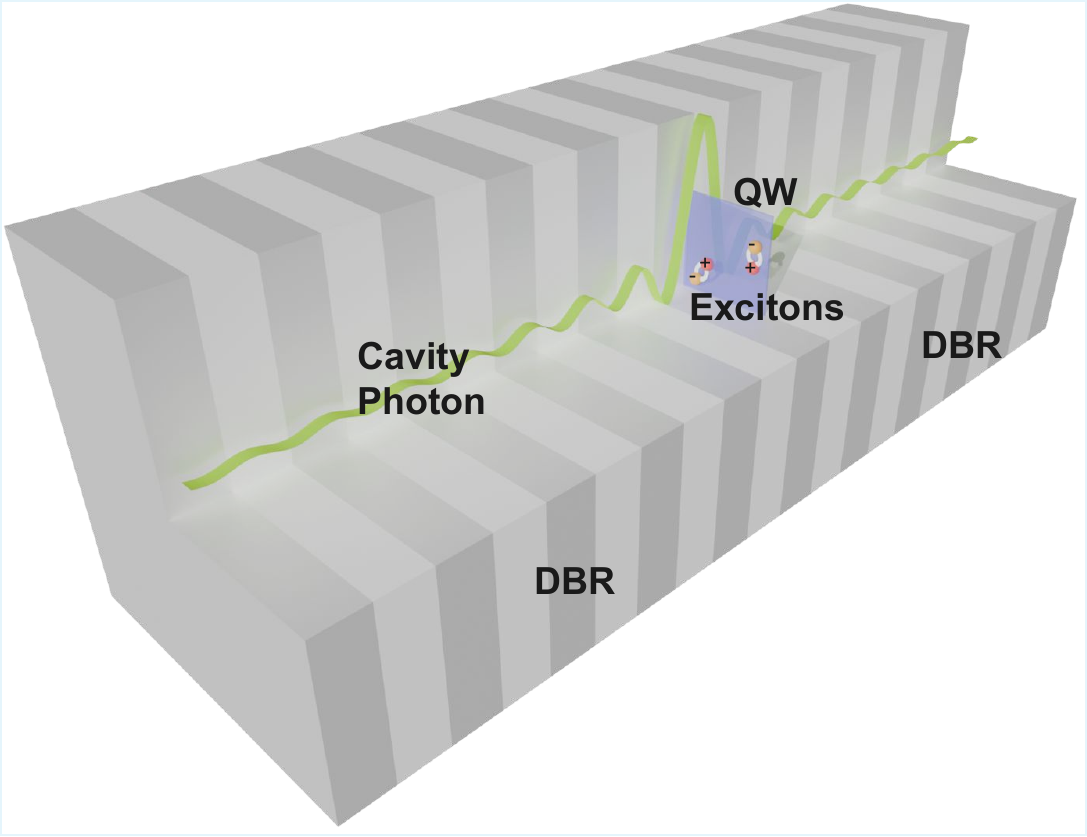}
    \caption{Scheme of a microcavity structure with two dielectric Bragg mirrors (DBR) confining a photon mode. The cavity incorporates an active quantum well (QW) region hosting excitons. The cavity structure is assumed to extend indefinitely in the cavity plane, while the active region is constricted to a finite volume.}
    \label{fig:schematics}
\end{figure}

Nevertheless, there are situations where descriptions based on Equation~\eqref{eq:Hopfield} and its generalizations fail. In the multimode case, when light-matter interaction energy is stronger than the bare energy level spacing, it modifies mode shapes so much that polariton modes can no longer be approximated as superpositions of a few bare modes. In the extreme case, an infinite number of bare modes may be needed to reconstruct the polariton mode - for example when polariton modes are localized while bare modes are plane waves. An example of such a system is a planar dielectric microcavity  that contains a finite volume of an active medium where \change{matter excitations} can form (see \textbf{Figure~\ref{fig:schematics}}). Other examples include non-orthogonal modes in nanophotonic structures and excitons bound by the interaction with light~\cite{cortese2022real,cortese2021excitons}.

A question arises whether it is possible to determine an accurate model of polaritons valid in an arbitrary physical structure that would be applicable in both classical and quantum regimes. Ideally, it would provide the form of a Hamiltonian (in the conservative case) or a quantum master equation (in the dissipative case), with coefficients resulting from material properties, without any fitting parameters. Moreover, the model should be efficient for analytical and numerical computations, meaning that the basis of modes should be optimal for the problem at hand. This issue is particularly important in the quantum regime where the sub-optimal choice of basis could lead to a massive increase of computational complexity.

Here, we propose a recipe for constructing a quantum mechanical model of polaritons in an arbitrary dielectric environment. We determine eigenmodes of a light-matter system starting from the Lagrangian density including light-matter interaction. In the conservative case, we employ Bogoliubov transformation to obtain the diagonal form of the Hamiltonian, with eigenmodes that are in correspondence to the solutions of classical Maxwell equations coupled to matter polarization. Our method supplements the treatments based on the approach of Huttner and Barnett~\cite{huttner1992quantization} by simplifying calculations and providing a straightforward quantum-classical correspondence. 
Next, we consider dissipative systems in the linear regime. While many theoretical approaches have been proposed in the past to describe photons interacting with the environment~\cite{huttner1992quantization,drummond1990electromagnetic,Santos,knoll1992quantum,milonni1995field,PhysRevA.53.1818,PhysRevA.57.4818,PhysRevA.57.3931,bechler1999quantum,PhysRevA.95.023831,raymer2020quantum,suttorp2004field,bhat2006hamiltonian}, none of them demonstrated a straightforward and general method to obtain a quantum master equation with no fitting parameters, valid in the strong coupling regime.
In this work, by means of the technique of third quantization~\cite{prosen2010quantization}, we find the optimal basis  that corresponds to normal superoperator modes of the dissipative system with linear losses. In analogy to the conservative case, we find that the solutions of Maxwell equations including  optical and material losses, as well as losses resulting from boundary conditions \change{(i.~e.~quasinormal modes~\cite{Ho_QNM,Lalanne_Review,Richter_QNM})}, correspond to the normal modes of the quantum Liouvillian. We show how this allows one to determine the exact form of the master equation in an optimal, diagonal basis.

As an application of our theory, we consider a nonlinear system of interacting \change{exciton-}polaritons. %We find that quantum corrections to the interactions lead to a geometry-dependent energy akin to the Casimir effect. 
We show how our theory can be used to engineer polariton many-body Hamiltonians. For instance, constricting the volume of emitters leads to the increase of two-body interactions between polaritons even in the absence of strong lateral optical confinement. This provides a new strategy in the quest for strongly interacting exciton-polaritons, which can boost existing methods that range from optical confinement~\cite{verger2006polariton,Munoz-Matutano2019,Delteil2019}, dipolar interactions~\cite{togan2018enhanced,datta2022highly,liran2024electrically} to polaron-mediated interactions~\cite{ravets2018polaron,emmanuele2020highly}. We also consider interaction between two spatially separated polariton modes and predict strong nonlocal interactions, which can be used to induce nonlocal correlations and mode entanglement of emitted light. \change{Moreover, our theory can be useful to model other polariton species, including phonon-polaritons, which exhibit both ultrastrong coupling and extreme confinement~\cite{galiffi2024extreme}.}

\section{Conservative system} \label{sec:conservative}

The problem of interaction of light and matter at the quantum level has been considered in great detail in the past~\cite{huttner1992quantization,drummond1990electromagnetic,Santos,knoll1992quantum,milonni1995field,PhysRevA.53.1818,PhysRevA.57.4818,PhysRevA.57.3931,bechler1999quantum,PhysRevA.95.023831,raymer2020quantum,suttorp2004field,bhat2006hamiltonian}, including the seminal work of Huttner and Barnett~\cite{huttner1992quantization}, which also treated interaction with matter reservoir modes. In this section, we propose an alternative approach to solving the diagonalization problem, which  simplifies the analysis and emphasizes a physically important aspect, the quantum-classical correspondence~\cite{schwennicke2024molecular}. Additionally, it provides a basis for the consideration of the dissipative case in the next section, ultimately leading to the formulation of the quantum master equation.

\subsection{Classical regime}

We consider coupling of the electromagnetic field and matter polarization under the assumption of the small size of emitters compared to the wavelength of light.
Lagrangian density of electromagnetic field in a dielectric medium coupled to localized matter modes can be written in the Coulomb gauge as $\mathcal{L}=\mathcal{L}_{E M}+\mathcal{L}_X+\mathcal{L}_{EM-X}$, with~\cite{Hopfield_Polaritons,GlauberLewenstein,huttner1992quantization} 
\begin{align}
& \mathcal{L}_{E M}=\frac{\epsilon_b({\bf r})}{2} \dot{\mathbf{A}}^2-\frac{1}{2 \mu_0}(\nabla \times \mathbf{A})^2, \\    
& \mathcal{L}_X=\frac{1}{2} \rho \dot{\mathbf{X}}^2-\frac{1}{2} \rho \omega_0^2 \mathbf{X}^2, \\
& \mathcal{L}_{E M-X}=-\alpha({\bf r}) \mathbf{A} \cdot \dot{\mathbf{X}},
\end{align}
where $\epsilon_b({\bf r})$ is the space-dependent background dielectric constant, $\rho$ is the mass density, $\omega_0$ is matter oscillator frequency, $\alpha(\er)$ is the light-matter coupling coefficient, $\mathbf{E}=-\dot{\mathbf{A}}$, and $\mathbf{B}=\nabla \times \mathbf{A}$. Polarization density can be identified as $\mathbf{P}=\alpha(\er)\mathbf{X}$.
The electromagnetic part of the Lagrangian corresponds to Maxwell equations in a dielectric.  The matter part  is described as harmonic oscillators. For simplicity, we assume that frequency and mass density of the matter component are space-independent, but these assumptions are not crucial for the following considerations. 
The background refractive index $\epsilon_b({\bf r})$ is here frequency independent, but later we will show how to incorporate dispersion of the background material. The above model is in some sense minimal, but can already precisely describe a broad range of inhomogeneous polariton systems. We will also see that generalizations to incorporate e.g.~anisotropy, magnetic response or electric currents are easy to implement.

While the above is written as a Lagrangian function in terms of continuous fields, we will actually have in mind a version of the model where a finite number of electromagnetic field modes and matter modes are considered. Physically, this corresponds to considering finite system volume $V$, with an appropriate UV cutoff for the electromagnetic field. For matter modes, the cutoff corresponds to the size of the emitter, such as an exciton or an atom. On the other hand, we assume that the spatial variation of the electromagnetic field component occurs on much larger spatial scales than the size of emitters, which justifies the use of the continuous  field to describe matter polarization. Despite such discretization of modes, we will use continuous integrals over volume $V$ for convenience, keeping in mind that they should be replaced by corresponding summations in the case of matter fields. We will not use continuity of these fields anywhere in the derivation. The total number of modes of the model will be denoted by $2N=2N_{\rm el}+2N_{\rm mat}$, where $2N_{\rm el}$ and $2N_{\rm mat}$ are the number of bare electromagnetic and matter modes, respectively, and the coefficient 2 stems from time-reversal symmetry. This discretization avoids certain computational issues occurring in the infinite-mode case. 

The Lagrangian $L=\int_V\mathcal{L}\,d\er$ corresponds to Maxwell equations coupled to polarization field. Their $2N$ solutions $(\mathbf{E}_j({\bf r})$, $\mathbf{X}_j({\bf r}))$ with $j=1\dots 2N$ obey
\begin{align}
\nabla \times \nabla \times \mathbf{E}_j=&\frac{\epsilon_b({\bf r})\omega_j^2}{\epsilon_0 c^2}  \mathbf{E}_j+ \frac{\alpha({\bf r})\omega_j^2}{\epsilon_0 c^2}  \mathbf{X}_j, \label{eq:Maxwell}\\
(\omega_0^2 - \omega_j^2) \label{eq:X} \mathbf{X}_j=&\frac{\alpha({\bf r})}{\rho} \mathbf{E}_j,
\end{align}

where $\omega_j$ is the oscillation frequency. We can move to the complex representation of the electromagnetic field where negative frequency eigenmodes are complex conjugates of their positive frequency partners. For each mode with frequency $\omega_j$, there exists a mode with frequency $-\omega_j$ and conjugated $E_j^*$ profile (i.e.~the magnetic field profile has a flipped sign). \change{In this representation, the matter field $\mathbf{X}_j$ becomes complex as well.}

The above equations can be recast as an eigenvalue problem for a Hermitian operator $\hat{O}$ on the composite six-dimensional vector $\mathbf{F} =\left(\mathbf{G}(\er),\mathbf{Y}(\er)\right)^T$ with $\mathbf{G}(\er)=\sqrt{\epsilon_b({\bf r})} \mathbf{E}(\er)$ and $\mathbf{Y}(\er)=\rho^{1/2} \omega_0\mathbf{X}(\er)$. Its eigenmodes form an orthogonal set
\begin{align}
\label{eq:normalization}
\int_V  {\bf F}^*_i(\er)  {\bf F}_j(\er)d \er &= \int_V  \left({\bf G}^*_i(\er) {\bf G}_j(\er) + {\bf Y}^*_i(\er) {\bf Y}_j(\er) \right)d \er=\nonumber\\&= \mathcal{N}_j\delta_{ij}
\end{align}
of solutions with real eigenvalues $\omega_j^2$, complete in a certain subspace. In particular, in the case $\epsilon_b(\er)=1$, $\alpha(\er)=0$ the subspace contains linear combinations of all matter modes and transversely polarized electromagnetic plane waves~\cite{GlauberLewenstein}. Here $\int_V d \er |{\bf G}_j(\er)|^2$ and $\int_V d \er |{\bf Y}_j(\er)|^2$ quantify the photonic and matter components of the polariton, in analogy to the squared Hopfield coeficients $|X|^2$ and $|C|^2$ in Equation~\eqref{eq:Hopfield}. Indeed, it can be verified that Equation\change{s}~\eqref{eq:Maxwell} and~\eqref{eq:X} transform to the Hopfield problem, Equation~\eqref{eq:Hopfield}, in the homogeneous case and under RWA $|\omega_j-\omega_0|\ll\omega_0$, provided that $\mathbf{G}=C\hat{e}\,e^{i\mathbf{k}\er}$, $\mathbf{Y}=X\hat{e}\,e^{i\mathbf{k}\er}$, $\Omega=-\hbar\alpha/\sqrt{\rho\varepsilon}$, $E_X=\hbar\omega_0$ and $E_C(k)=\hbar |\mathbf{k}|c\sqrt{\varepsilon_0/\varepsilon}$, where $\hat{e}$ is the polarization unit vector perpendicular to momentum $\mathbf{k}$.

The normalization coefficients $\mathcal{N}_j$ can be chosen to be unity, leading to an orthonormal set of eigenmodes. However, for reasons that will become clear later, we choose $\mathcal{N}_j$ in such a way that the energy of each mode (value of the classical Hamiltonian functional $H$) is equal to $\hbar |\omega_j|$. Hermiticity of operator $\hat{O}$ represented by Equation\change{s}~\eqref{eq:Maxwell} and~\eqref{eq:X} in the general case $\langle {\bf F}_1|\hat{O} {\bf F}_1\rangle=\langle \hat{O} {\bf F}_1|{\bf F}_2\rangle$ can be checked directly by integration by parts. We will assume that all of these classical solutions are stable, i.e.~$\omega_j^2>0$ for all $j$, which is a physically justified requirement of the stability of vacuum, as  energy cannot appear from nowhere. Note that this requirement is not fulfilled for all parameters, for example it is not the case when matter oscillators have a negative mass density $\rho$.

In the absence of coupling, $\alpha=0$,  Equation~\eqref{eq:Maxwell} describes electromagnetic field in a dielectric medium. In the presence of coupling, the solutions of the coupled equations~\eqref{eq:Maxwell} and~\eqref{eq:X} are polariton solutions in the classical regime. The corresponding Hamiltonian density $\mathcal{H}=\mathcal{H}_{E M}+\mathcal{H}_X+\mathcal{H}_{EM-X}$ and Hamiltonian $H=\int_V \mathcal{H}\,d \er$ are obtained by introducing canonical momenta $\Pi_{A,i}=\frac{\partial \mathcal{L}}{\partial  \dot{A}_i}=\epsilon_b(\mathbf{r}) \dot{A}_i$ (i.e.~$\mathbf{\Pi}_A=-\mathbf{D}$) and ${\Pi}_{X,i}=\frac{\partial\mathcal{L}}{\partial \dot{{X_i}}}=\alpha {A_i}+\rho \dot{{X_i}}\,$

\begin{align}
\mathcal{H}_{EM} =&\frac{\mathbf{\Pi}_A^2}{2\epsilon_b(\mathbf{r})}+\frac{1}{2\mu_0}(\mathbf{\nabla}\times \mathbf{A})^2\,,\\
 \mathcal{H}_{X}=&\frac{1}{2\rho}\mathbf{\Pi}_X^2+\frac{1}{2} \rho \omega_0^2 \mathbf{X}^2\,,\label{eq:isdhfioh4903hrfwoi}\\
 \mathcal{H}_{EM-X}=&-\frac{\alpha}{\rho}\mathbf{\Pi}_X \mathbf{A} + \frac{\alpha^2}{2\rho}\mathbf{A}^2\,,\label{eq:HEMX}
 \end{align}
 where $\mathcal{H}_{EM-X}$ includes the term $\frac{\alpha^2}{2\rho}\mathbf{A}^2$, which, in the quantum regime, corresponds to interaction of light with vacuum fluctuations of the matter field. 

\subsection{Quantization}
 
Canonical quantization of light and matter fields is performed by demanding appropriate commutation relations between variables and their conjugates. First, we determine the diagonal form of the above Hamiltonian. Instead of doing this explicitly, we will make use of known results. In the absence of light-matter coupling, $\alpha({\bf r})=0$, it is well known that both the electromagnetic field described by $\mathcal{H}_{EM}$ and the matter field described by $\mathcal{H}_{X}$ can be quantized as independent sets of harmonic oscillators. Assuming space of finite volume, the resulting diagonal form is $\hat{H}_{EM}=\sum_{+i} \hbar \omega_i \left(\hat{a}_i^\dagger \hat{a}_i+\frac{1}{2}\right)$~\cite{GlauberLewenstein}, where the 
 "$+i$" summation is over positive frequencies only, and in the Heisenberg picture the corresponding electromagnetic field and vector potential are 
\begin{align}
\hat{\bf E}({\bf r},t)&=\sum_{+i}  \hat{a}_i {\bf E}_i^{(\alpha=0)}({\bf r}) + {\rm h.c.}\\ \nonumber
\hat{\bf A}({\bf r},t)&=-i\sum_{+i}\frac{1}{\omega_i} %\left(\frac{\hbar}{2 \omega_i}\right)^{1/2}
 \hat{a}_i {\bf E}_i^{(\alpha=0)}({\bf r}) + {\rm h.c.}
 \end{align}
where ${\bf \change{E}}_i^{(\alpha=0)}({\bf r})$
 are eigensolutions to uncoupled Equation~(\ref{eq:Maxwell}). The difference with respect to~\cite{GlauberLewenstein} results from different choice of normalization coefficients $\mathcal{N}_i$. On the other hand, $\hat{H}_{X}=\sum_\nu\int_V d \change{\er}  \hbar \omega_0 \left(\hat{b}_\nu^\dagger(\change{\er}) \hat{b}_\nu(\change{\er})+\frac{1}{2}\right)$
 where $\nu=x,y,z$, while $\hat{a}_i$, $\hat{b}_\nu(\er)$ are annihilation operators obeying bosonic commutation relations.  Matter operator \[\hat{\bf X}({\bf r},t)=\left(\frac{\hbar}{2\rho\omega_0}\right)^{1/2} \sum_\nu \hat{b}_\nu({\bf r}) \vec{e}_\nu  + {\rm h.c.}\] is simply coordinate operator of a 3D harmonic oscillator at each (discrete) point in space where emitter mode exists. Since the total Hamiltonian is quadratic, conjugate momenta ${\bf \Pi}$ and ${\bf \Pi}_X$ can be written as linear combinations of operators $\hat{a}_i$, $\hat{a}_i^\dagger$, and $\hat{b}_\nu(\change{\er})$,  $\hat{b}^\dagger_\nu(\change{\er})$, respectively. \change{Note that here we are disregarding the possible non-bosonicity resulting from the internal structure of matter excitations, see eg.~\cite{kira2011semiconductor,szymanska2006nonequilibrium}.}

\subsection{Diagonalization in polariton modes}

Considering now the coupled case, $\alpha({\bf r})\neq0$, it is clear that the coupling Hamiltonian density $\mathcal{H}_{EM-X}$, Equation~(\ref{eq:HEMX}), after substitution of the above formulas, becomes quadratic in creation and annihilation modes of the \textit{uncoupled system}. Note that adding the coupling not only introduces a coupling term in the Hamiltonian but also alters the conjugate momentum corresponding to a particular coordinate, which is taken into account by the ${\bf A}^2$ term in Equation~(\ref{eq:HEMX}). This term, however, is also quadratic. The complete Hamiltonian has the general form $\hat{H}=\frac{1}{2}\sum_{ij}\hat{c}^\dagger_i \mathcal{A}_{ij} \hat{c}_j + \hat{c}^\dagger_i \mathcal{B}_{ij}\hat{c}^\dagger_j + {\rm h.c.}$, with $\mathcal{A},\mathcal{B}\in \mathbb{C}^{N\times N}$ where the set of operators $\hat{c}_i$ is composed of both $\hat{a}_i$ and $\hat{b}_\nu(\change{\er})$. Recall that $\hat{b}_\nu(\change{\er})$ includes a discrete set of emitter positions so this set is also finite. This Hamiltonian can be diagonalized using Bogoliubov transformation \[
\hat{c}_i = \sum_{+j} \left( u_{ij} \hat{P}_j + v_{ij}^* \hat{P}_j^\dagger \right)\]
into
\begin{equation}
    \label{eq:Hpol}
    \hat{H}=\sum_{+j}  \hbar \omega_j \left(\hat{P}_j^\dagger \hat{P}_j+\frac{1}{2}\right),
\end{equation}
where $\hat{P}_j$ are operators of polariton modes, which obey bosonic commutation relations due to symplecticity of the transformation. While the existence of such transformation is not guaranteed for bosons, it always exists when the eigenmodes of the associated Bogoliubov-de Gennes matrix $\hat{H}_{\text{BdG}}$ defined by
\begin{align}
\hat{H}+\mathrm{const}&= \frac{1}{2}\left(c^\dagger\,c\right)
\begin{pmatrix}
\mathcal{A} & \mathcal{B} \\
\mathcal{B}^\dagger & \mathcal{A}^T
\end{pmatrix} \begin{pmatrix}
c \\
c^\dagger
\end{pmatrix}
= \nonumber\\&= \frac{1}{2}\left(c^\dagger\,
c\right)\hat{H}_{\text{BdG}}
\begin{pmatrix}
c \\
c^\dagger
\end{pmatrix} 
\end{align}
are stable, which is equivalent to the positivity of eigenvalues or stability of the corresponding classical system~\cite{derezinski2017bosonic}. This condition is fulfilled whenever there is a well defined ground state of the system, i.e.~$\hat{H}_{\text{BdG}}$ is bounded from below. Due to the conservation of the number of modes under the transformation, the number of polariton modes in Equation~\eqref{eq:Hpol} is equal to the sum of bare electromagnetic and matter bosonic modes (equal to N since each bosonic mode corresponds to a pair of positive and negative frequency classical modes).

\subsection{Quantum-classical correspondence}

While the existence of Bogoliubov transformation does not give information about the explicit form of $\hat{P}_j$ operators, or coefficients $u_{ij},v^*_{ij}$, physical observable operators can be determined by consideration of the classical limit, i.e.~a coherent state corresponding to $j$-th polariton eigenmode $\change{\hat{P}}_j$, which we denote by $|\alpha_j\rangle$, where $\alpha_j$ is a complex number. Expectation values of electric field and matter coordinates have to obey Equation~(\ref{eq:Maxwell}) and (\ref{eq:X}). Since only quadratic terms are present in the Hamiltonian, this suggests that they are linear combinations of polariton operators of the form 
\begin{align}
\hat{\bf E}({\bf r},t)&=\sum_{+j}  \hat{P}_j {\bf E}_j({\bf r}) + {\rm h.c.}, \label{eq:Epol}\\
\hat{\bf X}({\bf r},t)&=\sum_{+j} \hat{P}_j {\bf X}_j({\bf r})  + {\rm h.c.} \label{eq:Xpol}
\end{align}
Since the Hamiltonian is diagonal in the basis $\hat{P}_j$, the first terms on the right hand sides are positive frequency parts $\hat{\bf E}^{(+)}$ and $\hat{\bf X}^{(+)}$, and their Hermitian conjugates are negative frequency parts $\hat{\bf E}^{(-)}$ and $\hat{\bf X}^{(-)}$. Normally ordered terms, where positive frequency operators are on the right of negative frequency operators, are to be used to determine physical observables \change{such as correlation functions}~\cite{scully1999quantum,GlauberLewenstein}. It is straightforward to check that  formulas~(\ref{eq:Epol}) and~(\ref{eq:Xpol}) give correct values of correlation functions in the classical limit. Consider the two-point first-order spatial correlation function
\begin{align}\label{eq:E1E2}
G^{(1)}_{\mu,\nu}&(\er_1,\er_2;t,t) =
\langle \alpha_j | { E}_\nu^{(-)}(\er_1,t) { E}_\mu^{(+)}(\er_2,t)| \alpha_j\rangle =\\ \nonumber&= |\alpha_j|^2 { E}_{j,\nu}^*(\er_1){ E}_{j,\mu}(\er_2) = \langle n_j  \rangle  { E}_{j,\nu}^*(\er_1){ E}_{j,\mu}(\er_2),
\end{align}
where $\mu,\nu=x,y,z$, and we used that for a coherent state  $\hat{P}_j|\alpha_j\rangle = \alpha_j |\alpha_j\rangle$. This is exactly the correlation function that one would expect for a coherent state in the classical limit, in the state given by ${\bf E}_j$ and ${\bf X}_j$, considering that ${\bf E}_j$ is the electric field corresponding to a coherent state with one polariton on average, $\langle n_j  \rangle=1$. Recall that ${\bf E}_j$ and ${\bf X}_j$ are normalized in Equation~\eqref{eq:normalization} such that the expectation value of the Hamiltonian is equal to $\hbar |\omega_j|$, and any real-valued classical electromagnetic field necessarily appears as an equal contribution of positive- and corresponding negative-frequency modes.

Analogously, correlations between ${\bf E}$ and ${\bf X}$ fields can be obtained. It is also clear that no  linear combinations of polariton operators other those given by Equation\change{s}~(\ref{eq:Epol}) and~(\ref{eq:Xpol}) can result in correct first-order correlation functions of all kinds, while nonlinear terms are not present in the Bogoliubov transformation.

The above relations can be inverted by multiplication of both sides (\ref{eq:Epol}) by $\epsilon_b(\er) {\bf E}_j^*({\bf r})$ and both sides of (\ref{eq:Xpol}) by $\rho \omega_0^2 {\bf X}_j^*({\bf r})$, adding to each other and integrating  over space to yield the explicit form of the polariton operator
\begin{equation}
    \hat{P}_j = \mathcal{N}_j^{-1}\int_V d{\bf r} \left(  \hat{\bf E} ({\bf r}) \epsilon_b(\er){\bf E}_j^*({\bf r}) + \hat{\bf X} ({\bf r}) \rho \omega_0^2 {\bf X}_j^*({\bf r})\right),
\end{equation}
where we have used orthogonality of $(\sqrt{\epsilon_b({\bf r})}{\bf E}(\er),\rho^{1/2} \omega_0 {\bf X}(\er))$. The polariton annihilation operator annihilates both photons and excitons according to the amplitudes $\epsilon_b(\er){\bf E}_j^*({\bf r})$ and $ \rho \omega_0^2 {\bf X}_j^*({\bf r})$, respectively.

In summary, an arbitrary linear light-matter system in the conservative case can be diagonalized using Bogoliubov transformation. Polariton eigenmodes are given explicitly in the electromagnetic field and matter polarization operators, Equation\change{s}~\eqref{eq:Epol} and~\eqref{eq:Xpol}, where the spatial shapes of the modes correspond to classical solutions of Maxwell equations coupled to the matter field.  We note that this result has been formulated in previous works, in particular in~\cite{huttner1992quantization} in the homogeneous case, and in~\cite{bhat2006hamiltonian} in the inhomogeneous case, which also included the effects of \change{frequency}-dependent dielectric constant. In our approach, explicit determination of the transformation and the corresponding coefficients is not required, since all the physically relevant operators can be obtained by consideration of the classical limit. Therefore, the \change{calculations} become much more concise. Importantly, as we demonstrate in the next section, this approach can be generalized to the dissipative case, which allows one to determine the form of the quantum master equation. 

\section{Dissipative system} \label{sec:dissipative}
In the dissipative case, we consider a quadratic system as described above, but coupled to reservoir modes, which results in loss/gain channels to exciton, photon, polariton, or other modes. In the Born-Markov approximation, reduced density matrix of the system follows Gorini–Kossakowski–Sudarshan–Lindblad (GKSL) master equation
\begin{equation}\label{eq:Lindblad}
\dot{\hat{\rho}} = \widehat{\mathcal{L}}\hat{\rho}  = \frac{1}{i\hbar} [\hat{H},\hat{\rho}] +\sum_\mu \gamma_\mu \left(\hat{L}_\mu \hat{\rho} \hat{L}_\mu^\dagger -\frac{1}{2}\left\{ \hat{L}_\mu^\dagger \hat{L}_\mu,\hat{\rho}\right\}\right),
\end{equation}
where $\mL$ is the Liouvillian superoperator (we will indicate superoperators using "wide hats"), $\hat{H}$ is the Hamiltonian part operator given by Equation~\eqref{eq:Hpol}, and $\mu$ enumerates different gain/loss channels with rates $\gamma_\mu$. In the polariton basis introduced in the previous section, these channels are defined as \[\hat{L}_\mu = \sum_j \left( l_{\mu,j} \hat{P}_j+g_{\mu,j} \hat{P}^\dagger_j\right),\] with arbitrary complex coefficients $l_{\mu,j}$ and $g_{\mu,j}$. %For simplicity, we assume that there are only decay processes, but extension to gain processes is straightforward.
We consider only linear Lindbladian terms to preserve the quadratic form of the Liouvillian, and assume that the system has a stable nonequilibrium steady state. In particular, in the absence of gain $g_{\mu,j}=0$, the steady state is the polariton vacuum which fulfills $P_j |0\rangle_P=0$. However, we will consider the general case, without assuming that the steady state is a pure state.

\subsection{Third quantization treatment} \label{sec:3Q}

In general, conservative and dissipative parts of~\eqref{eq:Lindblad} do not have the same diagonal basis of eigenmodes. To find a common basis and diagonalize the complete Liouvillian, one should perform a transformation that is an analog of Bogoliubov transformation, but in the dissipative case. Such a transformation for quadratic bosonic systems with linear Lindbladians \change{was} proposed under the name of third quantization~\cite{prosen2010quantization}. It turns out that while it is generally not possible to find a common basis in the operator space, it is possible to find an adequate superoperator basis. Note that third quantization can be implemented in several different bases~\cite{McDonald_3rdQ,Dupays_3rdQ}, each of them having certain advantages, but here we follow the original approach of~\cite{prosen2010quantization}, which will allow to use very similar arguments as in Sec.~\ref{sec:conservative}.

In third quantization, the central objects are superoperators, such as $\mL$, acting on operators such as the density matrix operator denoted as $|\hat{\rho}\rangle\!\rangle$. One can think of $|\hat{\rho}\rangle\!\rangle$ as a vectorized representation of the density matrix and superoperators as operators acting on these vectors. We are interested in expectation values of observables denoted by $(\!(\hat{A}|\hat{\rho}\rangle\!\rangle:=\Tr \,\hat{A}\hat{\rho}$, where $\hat{A}$ is an operator representing the observable, a member of dual space with respect to the space of density operators, in the sense that $\hat{A}\hat{\rho}$ products are trace-class~\cite{prosen2010quantization}. 
Basic superoperators are simply annihilation $\hat{a}_j$ or creation  $\hat{a}^\dagger_j$ operators acting from the left or from the right on the density matrix. We define superoperators $\wh{a}^\dL|\hat{\rho}\rangle\!\rangle = |\hat{a}^{\dagger}\hat{\rho}\rangle\!\rangle$ and $\wh{a}^\dR|\hat{\rho}\rangle\!\rangle = |\hat{\rho} \hat{a}^{\dagger}\rangle\!\rangle$, while $(\!(\hat{A}|\wh{a}^\dL = (\!(\hat{A}\hat{a}^{\dagger}|$ and $(\!(\hat{A}|\wh{a}^\dR = (\!(\hat{a}^{\dagger}\hat{A}|$ etc.
For each bosonic mode, one may define four "basic" superoperators
\begin{align}\label{eq:a01}
    \wh{a}_{0,j}&=\wh{a}_j^L, & \wh{a}'_{0,j}&=\wh{a}^\dL_j-\wh{a}^\dR_j, \\ \nonumber
    \wh{a}_{1,j}&=\wh{a}^\dR_j, & \wh{a}'_{1,j}&=\wh{a}^{R}_j-\wh{a}^{L}_j,
\end{align}
where $j=1..N$ is the bosonic mode index. It is straightforward to check that these superoperators satisfy commutation relations $[\wh{a}_{\mu,j},\wh{a}'_{\nu,k}]=\delta_{\mu,\nu}\delta_{j,k}$, $[\wh{a}_{\mu,j},\wh{a}_{\nu,k}]=[\wh{a}'_{\mu,j},\wh{a}'_{\nu,k}]=0$ with $\mu,\nu\in\{0,1\}$. One also has $\wh{a}_{\nu,j} |0\rangle\!\rangle=0$ where $|0\rangle\!\rangle$ is the density matrix of vacuum, and, using the cyclic property of trace for trace-class products, $(\!(1|\wh{a}'_{\nu,j}=0$, where $(\!(1|$ is the identity operator. This allows to define dual-Fock bases of density operators and observable operators
\begin{equation}\label{eq:Fock}
|\underline{m}\rangle\!\rangle = \prod_{\nu,j} \frac{(\wh{a}'_{\nu,j})^{m_{\nu,j}}}{\sqrt{m_{\nu,j}!}}\,|0\rangle\!\rangle,
\qquad
(\!( \underline{m}| 
= (\!( 1|\;\prod_{\nu,j} \frac{(\wh{a}_{\nu,j})^{m_{\nu,j}}}{\sqrt{m_{\nu,j}!}}.
\end{equation}
where the $2N$-component index $\underline{m}=(m_{\nu,j};\nu\in\{0,1\},j\in\{1..N\})^T$ and Fock states fulfill the biorthonormality condition $(\!(\underline{m}'|\underline{m}\rangle\!\rangle=\delta_{\underline{m}', \underline{m}}$. %Consequently, as long as linear combinations of $\underline{m}$ belong to the $\ell^2$ Hilbert space for both the operators $\hat{A}$ and the observables $\hat{\rho}$, products $\hat{A}\hat{\rho}$ are trace-class, as required. 
Note that elements of the dual-Fock basis are generally not density matrices nor physical observables, as they are not necessarily Hermitian or unit-trace. However, carefully chosen superpositions of dual-Fock basis vectors can be used to construct physical operators.

In~\cite{prosen2010quantization}, it was shown that quadratic Liouvillians of the form~\eqref{eq:Lindblad} generating stable and non-degenerate dynamics can be diagonalized as
\begin{equation} \label{eq:Lzeta}
\mL= -\sum_{r=1}^{2N} i \omega_r \wh{\zeta}'_r \wh{\zeta}_r,
\end{equation}
where $\omega_r$ can be interpreted as the complex mode frequency, while $\wh{\zeta}_r$ and $\wh{\zeta}'_r$ are {\it normal modes} superoperators given by
a symplectic transformation on the extended Nambu vector of superoperators $\wh{b}=(\wh{a}_{0,j},\wh{a}_{1,j},\wh{a}'_{0,j},\wh{a}'_{1,j})^T$
\begin{equation}
    \label{eq:Vb}(\wh{\zeta}_r,\wh{\zeta}'_r)^T =\mathbf{V} \wh{b},
\end{equation}
where $\mathbf{V}$ is a complex $4N\times4N$ matrix.
In general, $\wh{\zeta}_r$ and $\wh{\zeta}'_r$ are not Hermitian conjugate to each other, but due to symplecticity of the transformation, they satisfy commutation relations $[\wh{\zeta}_r,\wh{\zeta}'_s]=\delta_{r,s}$, $[\wh{\zeta}_r,\wh{\zeta}_s]=[\wh{\zeta}'_r,\wh{\zeta}'_s]=0$~\footnote{Therefore, $\wh{\zeta}_r$ together with $\wh{\zeta}'_r$ form a complete set in the 4N-dimensional space spanned by superoperators defined by Equation~\eqref{eq:a01}, and together with the bilinear form $\langle u,v\rangle_\zeta:= [u,v]$, a complex symplectic vector space.}. Moreover, $(\!(1|\wh{\zeta}'_r=0$, which ensures the conservation of trace in evolution under  Equation~\eqref{eq:Lzeta}. The form of \eqref{eq:Lzeta} guarantees that in the Fock space defined in Equation~\eqref{eq:Fock}, a stable system has a unique nonequilibrium steady state, the vacuum of $\wh{\zeta}_r$ superoperators, which we denote by $|0_\zeta\rangle\!\rangle:=|{\rm NESS}\rangle\!\rangle$.
Superoperators $\wh{\zeta}_r$ and $\wh{\zeta}'_r$ can be used, in analogy to $\wh{a}_{\mu,\nu}$ and $\wh{a}'_{\mu,\nu}$ in Equation~\eqref{eq:Fock}, to construct dual-Fock bases of density operators and observable operators by acting on $|{0_\zeta}\rangle\!\rangle$ and $(\!(1|$, respectively %In a subspace corresponding to a normal mode $r$, written as $|\hat{\rho}\rangle\!\rangle=\sum_m \sigma_m \frac{(\wh{\zeta}'_r)^m}{\sqrt{m!}}|0\rangle\!\rangle_\zeta$.
\begin{equation}\label{eq:FockZeta}
|\underline{m}\rangle\!\rangle_\zeta
=\prod_{r}\frac{\bigl(\wh{\zeta}'_{r}\bigr)^{m_{r}}}{\sqrt{m_{r}!}}\,
|0_\zeta\rangle\!\rangle,
\quad
(\!( \underline{m}| _\zeta
= (\!( 1|\;\prod_{r} \frac{(\wh{\zeta}_{r})^{m_{r}}}{\sqrt{m_{r}!}},
\end{equation}
where $\underline{m}=(m_{r};r\in\{1..2N\})^T$.

\subsection{Coherent states of dissipative modes}

Our approach is based on the quantum-classical correspondence. To make use of it, we define analogs of coherent states in the space of superoperators. Let us define $\wh{\zeta}^{+}_r$, $\wh{\zeta}'^{+}_r$ as superoperators that result in Hermitian conjugates with respect to $\wh{\zeta}_r$, $\wh{\zeta}'_r$, i.e.~$|\wh{\zeta}_r^{+}\hat{\rho}^\dagger\rangle\!\rangle=|(\wh{\zeta}_r \hat{\rho})^\dagger\rangle\!\rangle$ and~$|\wh{\zeta}_r'^{+}\hat{\rho}^\dagger\rangle\!\rangle=|(\wh{\zeta}_r' \hat{\rho})^\dagger\rangle\!\rangle$ (for all operators $\hat{\rho}$ from the Fock space defined via Equation~\eqref{eq:FockZeta}, even if $\hat{\rho}\neq\hat{\rho}^\dagger$)~\footnote{Considering $\zeta_r$, $\zeta'_r$ as operators acting on the Hilbert space of Hilbert-Schmidt operators, $\zeta^+_r$, $\zeta'^+_r$ are their respective Hermitian adjoints. However, this property will not be important for us, and in this work we do not call $\zeta^+_r$  Hermitian adjoints of $\zeta_r$ to avoid confusion. Both $\zeta_r$ and $\zeta^+_r$ play the role of annihilation operators while $\zeta'_r$ and $\zeta'^+_r$ the role of creation operators, as follows from Equation~\eqref{eq:FockZeta} and the commutation relations.}. It is easy to see from~\eqref{eq:a01} that these operators can be constructed by replacing $\wh{a}_{\nu,j}$ with $\wh{a}_{1-\nu,j}$. Taking Hermitian conjugate of Equation~\eqref{eq:Lzeta}, it follows from the conservation of Hermiticity during evolution that for every normal mode superoperator $\wh{\zeta}_r$ in~\eqref{eq:Lzeta} with complex frequency $\omega_r$, $\wh{\zeta}^{+}_r$ is a normal mode superoperator with frequency $-\omega^*_r$. Therefore, it is either another element of the set of normal modes in Equation~\eqref{eq:Lzeta}, or a linear combination of degenerate normal modes with the same frequency  $-\omega^*_r$ (our assumptions exclude $\omega_r=-\omega^*_r$). In the latter case, $\wh{\zeta}^{+}_r$ can be incorporated into the normal mode basis through  symplectic Gram-Schmidt procedure. Hence, operators $\wh{\zeta}^{+}_r$, $\wh{\zeta}'^{+}_r$ fulfill the commutation relations just as the other members of the set of normal modes, in particular~$[\wh{\zeta}'^{+}_r,\wh{\zeta}^{+}_s]=\delta_{r,s}$, $[\wh{\zeta}'^{+}_r,\wh{\zeta}_s]=[\wh{\zeta}^{+}_r,\wh{\zeta}_s]=[\wh{\zeta}'^{+}_r,\wh{\zeta}'_s]=0$. The Liouvillian can be rewritten as
\begin{equation} \label{eq:Lzeta_pp}
\mL= \sum_{+r}\left( -i \omega_r \wh{\zeta}'_r \wh{\zeta}_r +  i \omega_r^* \wh{\zeta}'^+_r \wh{\zeta}^+_r \right),
\end{equation}
where the summation is over $N$ normal modes with $\Re(\omega_r)>0$.

We define the coherent state corresponding to $r$-th normal mode as
\begin{equation} \label{eq:alpha}
|\alpha\rangle\!\rangle_r  = {\rm exp}\left(\alpha\wh{\zeta}'_r+\alpha^*\wh{\zeta}'^{+}_r\right)|0_\zeta\rangle\!\rangle,
\end{equation}
where $\alpha$ is a complex number. These operators are manifestly Hermitian, unlike, for instance, the  operator ${\rm e}^{\alpha\wh{\zeta}'_r -\alpha^*\wh{\zeta}_r}|0_\zeta\rangle\!\rangle$. Since $(\!(1|\wh{\zeta}'_r=(\!(1|\wh{\zeta}'^+_r=0$, we have $\Tr |\alpha\rangle\!\rangle_r = (\!(1|\alpha\rangle\!\rangle_r=1$. Note that these coherent states differ from these defined in~\cite{McDonald_3rdQ,Dupays_3rdQ}, which correspond to  the original basis of $\wh{a}_{\mu,j}$ operators and not to normal mode superoperators $\wh{\zeta}_r$. Using commutation relations, it is straightforward to show that 
\begin{equation} \label{eq:zeta_alpha}
    \wh{\zeta}_s |\alpha\rangle\!\rangle_r  = \alpha |\alpha\rangle\!\rangle_r \delta_{rs}, \qquad 
    \wh{\zeta}^{+}_s |\alpha\rangle\!\rangle_r  = \alpha^* |\alpha\rangle\!\rangle_r \delta_{rs}.
\end{equation}
Defining the "Heisenberg" time-dependent superoperators as $\wh{\zeta}_r(t):=\re^{-\mL t}\wh{\zeta}_r \,\re^{\mL t}$, we have $d\wh{\zeta}_r(t)/dt=[\wh{\zeta}_r(t),\mL ]$. Using commutation relations again, one can verify that $[\wh{\zeta}_r(t),\wh{\zeta}_s']=\delta_{r,s}\re^{ -i\omega_rt}$ and $\wh{\zeta}_r(t)=\wh{\zeta}_r\re^{-i\omega_r t}$. With this result at hand, we find that the evolution of coherent state operators fulfills
\begin{align} \label{eq:alphamotion}
\wh{\zeta}_r|\alpha_r(t)\rangle\!\rangle &= \wh{\zeta}_r \re^{\mL t} |\alpha\rangle\!\rangle_r  =  \re^{\mL t} \wh{\zeta}_r(t) |\alpha\rangle\!\rangle_r   =\\\nonumber &= \alpha  \re^{-i\omega_r t} |\alpha_r(t)\rangle\!\rangle,
\end{align}
and similarly  $\wh{\zeta}_r^+|\alpha_r(t)\rangle\!\rangle=\alpha^*  \re^{i\omega_r^* t}|\alpha_r(t)\rangle\!\rangle$, i.e.~coherent state operator remains coherent during evolution while the parameter $\alpha$ evolves with complex frequency $\omega_r$. 

Since $(\!(1|\wh{\zeta}'_r=0$, primed superoperators are linear superpositions of primed basic superoperators $\wh{a}'_{\mu,j}$ defined in Equation~\eqref{eq:a01}, i.e.~$\wh{\zeta}'_r=\sum_j(c_{r,j}\wh{a}_{0,j}'+d_{r,j}\wh{a}_{1,j}')$. In result, $|\alpha\rangle\!\rangle_r$ given by Equation~\eqref{eq:alpha} can be, in the case of a pure steady state $|0_\zeta\rangle\!\rangle=|0_\zeta\rangle\langle 0_\zeta|$, expressed as
%\begin{widetext}
\begin{align}
|\alpha\rangle\!\rangle_r =& \nonumber
\,\re^{\alpha\wh{\zeta}_r'+\alpha^*\wh{\zeta}_r'^+}|0_\zeta\rangle\!\rangle =  \re^{\sum_j \left( \alpha c_{r,j} \wh{a}_j^\dL- \alpha d_{r,j} \wh{a}_j^{ L}-\alpha^* c^*_{r,j} \wh{a}_j^{ L} +\alpha^* d^*_{r,j} \wh{a}_j^\dL\right)+ {\rm h.c.\!}^R}|0_\zeta\rangle\!\rangle = \\ \nonumber
= & \left(\prod_j \hat{D}_{r,j}( \alpha) \right)|0_\zeta\rangle\langle 0_\zeta| \prod_j \hat{D}_{r,j}^\dagger(\alpha) = \hat{D}_r(\alpha) |0_\zeta\rangle\langle 0_\zeta|\hat{D}_r^\dagger(\alpha),
\end{align}
%\end{widetext}
where "h.c.$\!^R$" is a Hermitian conjugate with operators acting from the right, $\hat{D}_{r,j}( \alpha) = \exp\left(\alpha \hat{a}^{\dagger}_{r,j} - \alpha^*\hat{a}_{r,j}\right)$ with $\hat{a}^\dagger_{r,j}=c_{r,j} \hat{a}_j^{\dagger}-  d_{r,j} \hat{a}_j$, and $\hat{D}_r(\alpha)=\prod_{\change{j}} \hat{D}_{r,j}(\alpha)$ is the displacement operator corresponding to normal mode $r$, defined as
\begin{equation}
    \label{eq:ar}
\hat{a}_r^\dagger=\sum_j (c_{r,j}\hat{a}_j^\dagger \change{-} d_{r,j}\hat{a}_j), 
\end{equation}
In the case when the steady state is a mixed state, $|0_\zeta\rangle\!\rangle=\sum_np_n|n_\zeta\rangle\langle n_\zeta|$, we have\\ $|\alpha\rangle\!\rangle_r = \sum_n p_n \hat{D}_r(\alpha)|n_\zeta\rangle\langle n_\zeta| \hat{D}^\dagger_r(\alpha)$. We conclude that the dissipative coherent state operator $|\alpha\rangle\!\rangle_r$ is positive semi-definite, and it corresponds to a physical state.

Clearly, the dissipative coherent state is a displaced nonequilibrium steady state, and it remains so during evolution. This generalizes the known result that pure coherent states remain pure and coherent during single-mode evolution with linear decay~\cite{Gerry_Knight_2004}. The operators $\hat{a}_r$ corresponding to different normal modes do not commute with each other in general, $[\hat{a}^\dagger_r,\hat{a}_s]\neq\delta_{r,s}$,  since the classical solutions are not necessarily orthogonal in a non-Hermitian system. 
We emphasize that these coherent states are not necessarily pure states since the nonequilibrium steady state $|0_\zeta\rangle\!\rangle$ can be a mixed state. Nevertheless, the measured correlation functions exhibit perfect first-order coherence in the large particle number limit, as will be demonstrated below.

\subsection{Quantum-classical correspondence}

In analogy to the conservative case, we will demonstrate the correspondence of \change{the above defined} coherent states to classical states obeying Maxwell equations, but this time including the effects of dissipation.
As shown above, $N$ bosonic modes give rise to $2N$ normal modes of the Liouvillian. On the other hand, a single bosonic mode corresponds in the classical limit to a pair of Maxwell equations eigenmodes with positive and negative frequency. Hence, the $2N$ eigenmodes of Maxwell equations have $2N$ corresponding normal modes of the Liouvillian. 
To demonstrate direct quantum-classical correspondence, we must find the form of the electric field and matter operators in normal modes and compare them with solutions in the classical limit. 

As described in Sec.~\ref{sec:3Q}, in third quantization operators of observables are constructed by acting from the right side on the identity operator $(\!(1|$, see Equation~\eqref{eq:FockZeta}. Since electric field intensity and matter particle density are linear in the total particle number, their superoperators must include  linear terms in $\wh{\zeta}_r$ only. In analogy to the Hermitian case, Equation~\eqref{eq:Epol} and~\eqref{eq:Xpol}, we propose that
\begin{align}
(\!(\hat{\bf E}({\bf r})|&=(\!(1|\wh{\bf E}({\bf r})=(\!(1|\sum_r \left(\wh{\zeta}_r {\bf E}_r({\bf r}) + \wh{\zeta}^{+}_r {\bf E}^*_r({\bf r})\right), \label{eq:Ezeta}\\
(\!(\hat{\bf X}({\bf r})|&=(\!(1|\wh{\bf X}({\bf r})=(\!(1|\sum_r  \left(\wh{\zeta}_r {\bf X}_r({\bf r}) + \wh{\zeta}^{+}_r {\bf X}^*_r({\bf r})\right).  \label{eq:Xzeta}
\end{align}
The first and the second terms on the right hand sides correspond to positive and negative frequency parts, respectively. 
To check if this guess is correct, let us assume that the state of the system corresponds to the single-mode coherent state $|\alpha\rangle\!\rangle_r $, and calculate arbitrary first-order correlation functions, in analogy to Equation~\eqref{eq:E1E2}
\begin{align}\label{eq:E1E2zeta} \nonumber
 G_{\mu,\nu}^{(1)}&(\er_1,\er_2;t,t) =
(\!({ E}_\nu^{L(-)}(\er_1,t) { E}_\mu^{L(+)}(\er_2,t)| \alpha\rangle\!\rangle_r =\\ &
= (\!( { E}_{r,\nu}^*(\er_1){ E}_{r,\mu}(\er_2) \wh{\zeta}^{+}_r\wh{\zeta}_r| \alpha\rangle\!\rangle_r + G^{(1)}_{NESS} = \\ \nonumber &
=|\alpha|^2 { E}_{r,\nu}^*(\er_1){ E}_{r,\mu}(\er_2)  + G^{(1)}_{NESS} \\\nonumber &= \langle n  \rangle { E}_{r,\nu}^*(\er_1){ E}_{r,\mu}(\er_2) + G^{(1)}_{NESS},
\end{align}
where $G^{(1)}_{NESS}$ is the correlation function $G_{\mu,\nu}^{(1)}(\er_1,\er_2;t,t)$ in the nonequilibrium steady state $| 0_\zeta\rangle\!\rangle$. This term appears since the electric field superoperator $\wh{\bf E}$ may contain also primed $\wh{\zeta}'_r$ terms, which does not contradict Equation~\eqref{eq:Ezeta} since $(\!(1|\wh{\zeta}'_r=0$.  Nevertheless, in the classical limit we obtain first-order correlation functions that correspond exactly to expected correlations in solutions of Maxwell equations. Clearly, no other choice of space-dependent coefficients in Equation~\eqref{eq:Ezeta} and~\eqref{eq:Xzeta} can lead to the correct result.

Consequently, coherent states of normal modes $|\alpha\rangle\!\rangle_r$ of the above system can be assigned to solutions of Maxwell equations coupled to matter modes, Equation~(\ref{eq:Maxwell}) and (\ref{eq:X}), but with the addition of decay/gain channels. Possible channels include matter decay $\gamma_x$ and photon decay described with the imaginary part of the background refractive index $\epsilon_{bI}(\er)$. The corresponding system of eigenmode equations reads
\begin{align}
\nabla \times \nabla \times \mathbf{E}_r&=\frac{\left(\epsilon_{b\change{R}}({\bf r})+i\epsilon_{bI}({\bf r})\right)\omega_r^2}{\epsilon_0 c^2}  \mathbf{E}_r+ \frac{\alpha({\bf r})\omega_r^2}{\epsilon_0 c^2}  \mathbf{X}_r, \nonumber\\ \label{eq:MaxwellD}
(\omega_0^2 - 2 i \gamma_x \omega_r &-\omega_r^2) \mathbf{X}_r=\frac{\alpha({\bf r})}{\rho} \mathbf{E}_r.
\end{align}

Moreover, we can include decay resulting from outgoing-wave boundary conditions. In this case, eigenmodes of wave equations such as Equation~\eqref{eq:MaxwellD} are known as quasinormal modes (QNMs), and have been successfully used to describe systems ranging from black holes to subwavelength plasmonic resonators~\change{\cite{Lalanne_Review,Hughes_QNM,Richter_QNM,Ho_QNM,ren2022connecting}.} The description of light in terms of QNMs is a long-standing problem that has seen much controversy and debate, in particular concerning their normalization~\cite{Kristensen_Normalization,Langbein_comment}. These issues now \change{appear to be resolved}, with \change{a} clear definition of regimes where different existing normalization schemes are applicable, and equivalent~\cite{Lalanne_Review}. The difficulty  \change{in} using QNMs lies in the fact that they are divergent at $|\er| \to \infty$, which is a direct consequence of natural temporal divergence at $t \to -\infty$~\cite{Lalanne_Reflections}. However, we can make use of the regularization developed in classical systems to cope with this divergence. Assuming that our system of interest is localized in space, we can surround it with an absorbing material, for example an equivalent of a uniaxial perfectly matched layer (PML), which is a layer of anisotropic material designed to absorb any outgoing waves~\cite{UPML}. This leads to regularized QNMs of Maxwell equations which are exactly equivalent to QNMs of the original problem within the region of interest, do not suffer from the divergence problem, and can be properly normalized~\cite{Lalanne_Review}. Note that a simple isotropic lossy material would also suffice to absorb outgoing waves, as long as it is "switched on" sufficiently slowly as one moves to $|\er| \to \infty$.

The second equation of the system~\eqref{eq:MaxwellD} can be incorporated in the first equation as a frequency-dependent contribution to the refractive index. This equation can be then solved analytically or numerically using a Maxwell equation solver that can find eigenmodes in frequency domain with frequency-dependent complex dielectric constant and outgoing wave boundary conditions. Calculated eigenmodes and complex eigenvalues correspond to normal modes $r$ of the Liouvillian and their complex frequencies via Equation~\eqref{eq:Ezeta} and~\eqref{eq:Xzeta}. 

In fact, this equivalence of frequency-dependent refractive index and coupling to additional modes is the basis of "auxiliary modes" technique for formulating dispersive Maxwell equations as a non-dispersive  problem~\cite{Fan_Auxiliary,Lalanne_Rigorous}. Likewise, in our case we can model dispersive background material by auxiliary quantum matter modes which are weakly coupled to light. This way one can introduce frequency-dependent refractive index of the dielectric, in a way similar to that presented in~\cite{huttner1992quantization,bhat2006hamiltonian}. 

\subsection{Lindbladian form of master equation} \label{sec:Lindbladian}

We can transform the Liouvillian~\eqref{eq:Lzeta_pp} into a more familiar Lindbladian-like form that does not involve superoperators on the right hand side. This final step leads to an important practical result. The description of a bosonic system in the linear regime can be performed exactly using a set of independent (diagonal) single-mode Liouvillians, each of them corresponding to a single QNM, see~\eqref{eq:master}. Therefore, we directly indicate the optimal basis and the corresponding model. Since in practice quantum systems are usually described using only a small number of basis vectors for the sake of computational efficiency, the use of the optimal basis can lead to substantial improvements of both accuracy and complexity of the description.

Let us consider an arbitrary set of $N$ bosonic operators satisfying canonical commutation relations $[\hat{b}_r,\hat{b}^\dagger_s]=\delta_{rs}$, $[\hat{b}^\dagger_r,\hat{b}^\dagger_s]=[\hat{b}_r,\hat{b}_s]=0,$ with a common vacuum $\ket{0_b}$. We define a new set of superoperators 
\begin{align}\label{eq:eta}
&\wh{\eta}_r=\wh{b}_r^L,&  &\wh{\eta}'_r=\wh{b}^\dL_r-\wh{b}^\dR_r, \\\nonumber
&\wh{\eta}^+_r=\wh{b}_r^\dR,&  &\wh{\eta}'^+_r=\wh{b}^R_r-\wh{b}^L_r, \end{align}
cf.~Equation~\eqref{eq:a01}.
It is easy to see that 
$[\wh{\eta}_r,\wh{\eta}_s'] = \delta_{rs},
[\wh{\eta}'_r,\wh{\eta}'_s] = [\wh{\eta}_r,\wh{\eta}_s] = 0$.
The two Fock bases 
\begin{align}\label{eq:twinFock}
|\underline{m}\rangle\!\rangle_\zeta
=&\prod_{r}\frac{\bigl(\wh{\zeta}'_{r}\bigr)^{m_{0,r}}}{\sqrt{m_{0,r}!}}
\prod_{r}\frac{\bigl(\wh{\zeta}'^+_{r}\bigr)^{m_{1,r}}}{\sqrt{m_{1,r}!}}\,
|0_\zeta\rangle\!\rangle,
\\\nonumber
|\underline{m}\rangle\!\rangle_{{\eta}}
=&\prod_{r}\frac{\bigl(\wh{\eta}'_{r}\bigr)^{m_{0,r}}}{\sqrt{m_{0,r}!}}
\prod_{r}\frac{\bigl(\wh{\eta}'^+_{r}\bigr)^{m_{1,r}}}{\sqrt{m_{1,r}!}}\,
|0_b\rangle\!\rangle,
\end{align}
define two Fock spaces $\mathcal{F}_\zeta$ and $\mathcal{F}_\eta$, where the $2N$-component index $\underline{m}=(m_{\nu,r};\nu\in\{0,1\},r\in\{1..N\})^T$. Note that $\mathcal{F}_\zeta$ is identical to the one defined in Equation~\eqref{eq:FockZeta}. 
The  linear map
$f$ between $\mathcal{F}_\zeta$ and $\mathcal{F}_\eta$ such that $f(\kett{ \underline{m}}_\zeta)=\kett{ \underline{m}}_{{\eta}}$ is an isomorphism with respect to the action of superoperators. 
Consider now the following "twin Liouvillian" acting on operators from the space spanned by $|\underline{m}\rangle\!\rangle_{\eta}$
\begin{equation}\label{eq:tildeL}
\mL^b
= \sum_{+r}\Bigl(-i\omega_r{\wh{\eta}}_r'\wh{\eta}_r
+i\omega_r^*{\wh{\eta}}_r'^+\wh{\eta}_r^+\Bigr),
\end{equation}
cf.~Equation~\eqref{eq:Lzeta_pp}. Considering that 
\begin{align}
&\wh{\zeta}_r\,
|0_\zeta\rangle\!\rangle = 0, \quad
&\wh{\zeta}_r^+\, |0_\zeta\rangle\!\rangle &= 0, \\\nonumber
&\wh{\eta}_r\,
|0_b\rangle\!\rangle =  \wh{b}_r^L\,|0_b\rangle\!\rangle = 0,
\quad
&\wh{\eta}^+_r\,
|0_b\rangle\!\rangle &=  \wh{b}_r^\dR\,|0_b\rangle\!\rangle = 0,
\end{align}
and that both sets of superoperators fulfill the same commutation relations,
it follows that if $\hat{\rho}_\zeta$ and  $\hat{\rho}_\eta$ are operators from Fock spaces $\mathcal{F}_\zeta$ and $\mathcal{F}_\eta$, respectively,  such that $\hat{\rho}_\eta=f(\hat{\rho}_\zeta)$, then  $\mL^b\hat{\rho}_\eta=f(\mL \hat{\rho}_\zeta)$. Therefore, $\mL^b$ is equivalent conjugate (similar operator) to $\mL $ via isomorphism $f$. In other words, matrix representation of $\mL $ in the $\kett{ \underline{m}}_{\zeta}$ basis is identical to matrix representation of $\mL^b$ in the $\kett{ \underline{m}}_{\eta}$ basis. Moreover, the isomorphic mapping between $\hat{\rho}_\eta$ and $\hat{\rho}_\zeta$ is conserved in time if $\hat{\rho}_\eta(t)$ evolves according to~\eqref{eq:tildeL} and $\hat{\rho}_\zeta(t)$ evolves according to~\eqref{eq:Lzeta_pp}. 

Now, we can rewrite $\mL^b$ superoperator in a diagonal form using Equation\eqref{eq:eta} as %The advantage of considering $\mL^b$ instead of $\mL $ is that it is a sum  
$\mL^b= \sum_{+r}\mL^b_r$, where $\mL^b_r$ contains $\hat{b}_r$ and $\hat{b}^\dagger_r$ operators only
\begin{align}\label{eq:master_eta}
\mL^b_{r}\hat{\rho}_\eta
=& \left(-i\omega_r{\wh{\eta}}_r'\wh{\eta}_r
+i\omega_r^*{\wh{\eta}}_r'{}^+\wh{\eta}_r^+\right)\hat{\rho}_\eta=\\ \nonumber
=&\left(-i\omega_r(\wh{b}_r^{\dagger L }-\wh{b}_r^{\dagger R })\wh{b}_r^L
   + i\omega_r^*(\wh{b}_r^R - \wh{b}_r^L)\wh{b}_r^\dR\right)\hat{\rho}_\eta=\\ \nonumber
   =& -i\omega_r(\hat{b}_r^\dagger \hat{b}_r\hat{\rho}_\eta - \hat{b}_r\hat{\rho}_\eta \hat{b}_r^\dagger)
   + i\omega_r^*(\hat{\rho}_\eta \hat{b}_r^\dagger \hat{b}_r - \hat{b}_r\hat{\rho}_\eta \hat{b}_r^\dagger),
\end{align}
which leads to
%\begin{widetext}
\begin{align} \label{eq:master}
\mL _{b}\hat{\rho}
=\sum_r\Bigl(i\Re(\omega_r)\bigl[\hat{\rho},\hat{b}_r^\dagger \hat{b}_r\bigr]-\Im(\omega_r)\bigl(2 \hat{b}_r\hat{\rho} \hat{b}_r^\dagger  - \{\hat{b}_r^\dagger \hat{b}_r,\hat{\rho}\}\bigr)\Bigr).
\end{align}
%\end{widetext}
Where we omitted the index $\eta$. In result, evolution governed by a quadratic Liouvillian is isomorphic to evolution under a set of independent single-mode quadratic Liouvillians of the GKLS form. The description of evolution with the Lindbladian equation in the non-Hermitian normal basis is thus just as simple as Lindbladian evolution in the Hermitian basis, cf.~Equation~\eqref{eq:Lindblad}, but has the important advantage of having a diagonal form. Consequently, one can describe the evolution of each mode of the system exactly using the well-known single-mode \change{Lindbladian} description, without introducing any error due to the finite size of the basis. In particular, when only a few modes of the system are important (e.g.~due to energy considerations under resonant excitation), one can focus on the selected modes without worrying about the non-diagonal terms and the strength of coupling to other modes. That said, one must take care of the nontrivial nature of the vacuum state $| 0_{{\zeta}}\rangle\!\rangle$ in Equation~\eqref{eq:twinFock} and the isomorphism between superoperators when considering specific initial conditions, coupling the system to external modes, or adding non-quadratic terms.

We emphasize that our approach goes beyond the previous attempts of quantization in the basis of quasinormal modes~\cite{Richter_QNM,Hughes_QNM}, where the resulting Liovillian contained mode-mixing terms, and was therefore non-separable. In our case, both the Hamiltonian and the Lindbladian part of the master equation are obtained in the same diagonal basis, which is the crucial feature allowing for mode separation. It comes at the price of considering an isomorphic representation rather than the original one, but this has no influence on the form of the resulting master equation.

\section{\change{Why diagonalize?}}

We showed that in analogy to the conservative case, solutions of Maxwell equations coupled to matter form a basis of normal modes of the Liouvillian superoperator. This results provides a simple recipe for determining the \change{diagonal form of} quantum master equation for a linear system, which consists of: (1) Determining classical solutions of Maxwell equations coupled to matter with appropriate boundary conditions (QNMs), typically by incorporating the response of resonant matter modes into the complex refractive index; (2) Extracting the  frequencies with positive real parts and the corresponding eigenmodes and using them as a basis for master equation of the form~\eqref{eq:master}. 
%The first step may involve calculating the profile of ${\bf X}_i(r)$ from ${\bf E}_i(r)$ using the second equation of~\eqref{eq:MaxwellD} and the determined eigenfrequencies $\omega_i$.

\change{Still, it may remain unclear for the reader what are the advantages of using the above diagonalized description to model physical systems, rather than well-established methods, such as the standard GKLS description of the form~\eqref{eq:Lindblad}. }

\change{Let us first note that including even weak dissipation can dramatically affect the modes and system dynamics even on the classical level. For example, including decay will usually change the spatial shape of the modes, and in general lead to “density currents” within the modes, which are absent in the conservative case. This is exactly the case of QNMs, which often “leak out” of the confined high-intensity region, but it can also have the form of a density flow between light and matter components or various other non-Hermitian effects.}

\change{A typical quantization procedure may consist of using a Hermitian (orthogonal) basis to describe a dissipative system (where eigenmodes are non-orthogonal). One can ask a question, will there be any corrections to the predictions if we use the dissipative basis instead, as described in our work? Are there any measurable corrections or extra terms that can be identified?}

\change{In principle, there are no tangible "corrections" to the existing theory, because we know very well how to quantize electromagnetic field and matter, and actually it can be performed, at least formally, in many ways, without approximations. Our theoretical description does not make the existing fundamentals incorrect.}

\change{In practice, however, corrections can appear, but they do not come from specific new terms. Instead, they result from the way the "usual treatment" is performed in practice. The main issue when describing multimode quantum systems is the size of the Hilbert space. One typically is forced to choose a small basis of modes to deal with the exponential growth of complexity. These are to some extent arbitrary choices. Often, a basis of eigenstates of uncoupled modes is chosen to describe the light field, the matter field, and then some coupling is added. Nearly always, a basis corresponding to a non-dissipative model is used to describe a dissipative system. However, there is no guarantee that these will be good choices. Because one cannot use a large basis to describe a quantum system, this will always introduce some error, usually not a small one. This is the case even if we know that only a few modes contribute to system dynamics, because, for example, they are within a relevant energy range in a resonant experiment. Generally, coupling or dissipation will bring admixtures to neglected modes, substantially changing the eigenstates. For example, even a small leakage from a cavity may drastically change the shape of a mode that was a simple standing wave in the conservative case. This usually results in an uncontrollable error.  A more refined approach is to use QNM basis for a quantum model, as in Refs.~\cite{Richter_QNM,Ho_QNM}, but even then, coupling between modes exists because of the non-diagonal form of the master equation, which means that there will be an error when we use a finite basis. }

\change{On the other hand, in our approach we find an optimal basis in the superoperator space to describe a linear system. This leads to an exact description of the system for each mode. We know that there is no coupling between normal modes, so we can safely describe dynamics within particular modes, without worrying that a finite basis effect will introduce an error. When investigating a particular mode, we do not need to keep track of intermode correlations, so we can use a much smaller Hilbert space basis. In this sense, our approach indeed can lead to better theoretical predictions, since we are able to model a system avoiding a systematic error, and at the same time, keep the computation cost at relatively low level.}

\change{In some sense, the transformation in superoperator space considered in our work is analogous to a unitary transformation in quantum mechanics. A well designed unitary transformation does not bring any corrections to system description, but can make theoretical description much simpler or even tractable. In many cases, it can even improve our understanding of the system. Third quantization transformation is a specific generalization of Bogoliubov transformation, which is itself a unitary transformation, expressed in the second quantized form~\cite{GlauberLewenstein}.}

\change{Finally, we note that in the case of many typical systems, there is no need to consider the theory of third quantization to use the results of this work, since the resulting Eq.~\eqref{eq:master} has a standard Lindbladian form. This approach will be exemplified in the next section.}

\section{Applications}

Transformations introduced in previous sections focused on the determination of a diagonal basis in a quadratic system. While they resulted in \change{an} \textit{optimal} basis in quadratic systems, they can still result in a \textit{convenient} basis in a non-quadratic system. One may draw analogy to the convenient plane-wave basis which is commonly used to analyze systems in which plane waves are not eigenstates of the Hamilonian. Below, we will demonstrate in several practical examples how our approach can be useful in non-quadratic, interacting systems.

\subsection{Interactions}

For the interaction Hamiltonian, we consider
\begin{equation}\label{eq:X4}
\mathcal{H}_{\rm int}=\frac{g}{6} {\bf \hat{X}}^4.
\end{equation}
Substituting Equation~(\ref{eq:Xpol}), we obtain under the RWA
\begin{equation}\label{eq:Hint}
\hat{H}_{\rm int}=g \sum_{jl} \hat{P}^\dagger_l\hat{P}^\dagger_j\hat{P}_j\hat{P}_l \int_V d\er |{\bf X}_j(\er)|^2|{\bf X}_l(\er)|^2 + O(n).
\end{equation}
The coefficient $g$ has now a clear physical interpretation and is simply the interaction constant of the matter component, which corresponds to interaction strength of polaritons when the matter Hopfield coefficient is close to unity. General interaction Hamiltonians with other forms than Equation~\eqref{eq:X4} may be treated in the same way. Note that we have not made any approximation regarding the weakness of the nonlinearity of the system, and our treatment is still exact. However, the polariton basis $\hat{P}_{\change{j}}$ will be useful in practice mainly in the case when one is able to identify a small subset of polariton eigenmodes that are important. This is mostly the case when interactions are weak compared to other energy scales in the system.

\subsection{Quantum fluctuations}

We consider a single polariton mode, and take into account the lower-order terms that were neglected in the previous subsection (we stay within the RWA)
\begin{equation}
\hat{H}_{\rm int}=g \int_V |{\bf X}(\er)|^4 d\er \left(\hat{P}^\dagger\hat{P}^\dagger\hat{P}\hat{P} + 2 \hat{P}^\dagger\hat{P} + \frac{1}{2}\right).
\end{equation}
The two last terms in \change{parentheses} can be interpreted as the interaction between non-vacuum polaritons and vacuum fluctuations, and the interactions within the polariton vacuum, respectively. These terms can give rise to physically measurable effects. 
For instance, the second term leads to interaction-induced shift of the energy of polariton excitations. This can be observed in spectroscopic measurements as a blueshift of the polariton energy in the linear regime, $\hbar\omega_i^{\rm int}=\hbar\omega_i+2g\int|{\bf X}(\er)|^4 d\er$,
where $\omega_i^{\rm int}$ is the low-density polariton excitation frequency including the effect of interactions. The magnitude of this energy shift is comparable to the single-polariton nonlinearity. It is density-independent, but proportional to the {\it square} of the exciton fraction, in contrast to other density-independent effects which are simply proportional to the exciton fraction.

\subsection{Exciton confinement}

\begin{figure*}
    \centering    \includegraphics[width=0.9\linewidth]{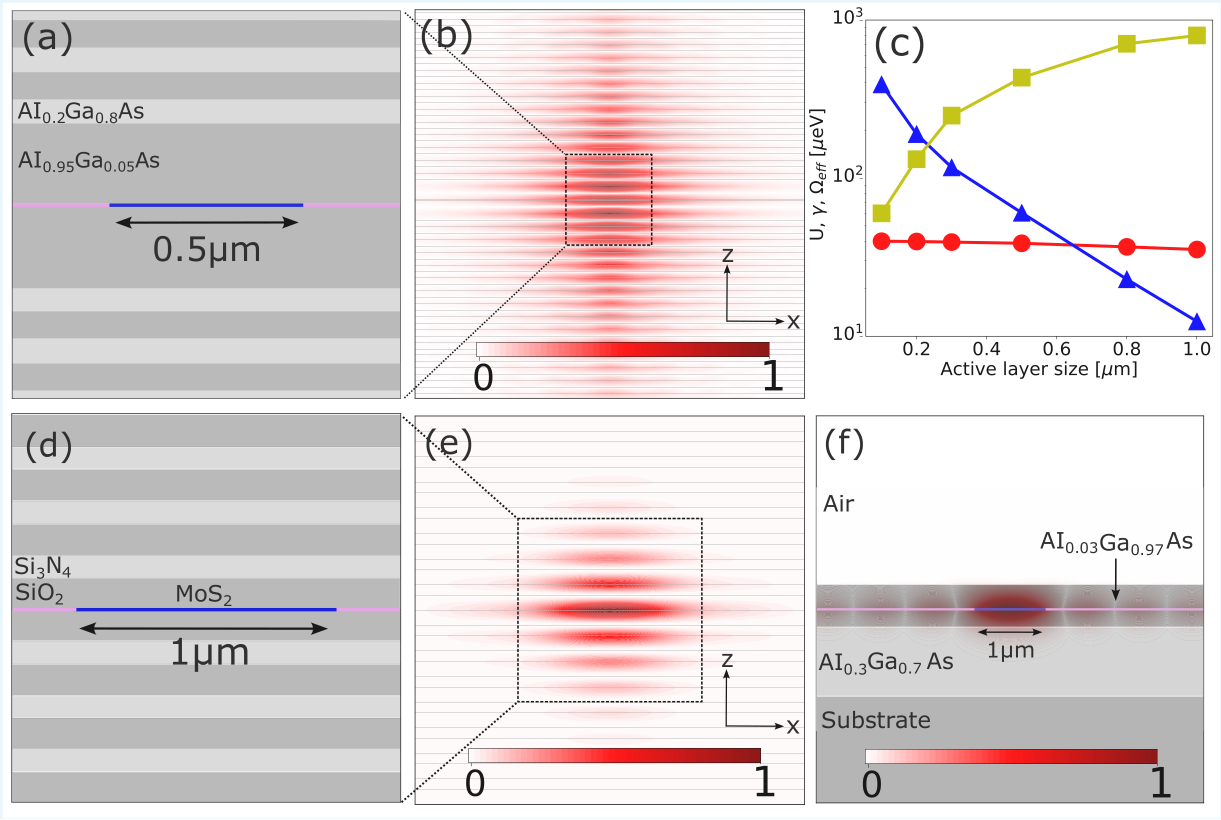}
    \caption{(a) Proposed vertical microcavity structure with  no structural lateral confinement for photons. Blue area corresponds to InGaAs active layer with excitons resonant with the optical mode, and pink area is the Al$_{0.95}$Ga$_{0.05}$As layer with no resonant exciton transition. (b) Light intensity in the lowest energy polariton mode for the structure shown in panel (a). (c) Calculated polariton interaction U (blue triangles), loss rate $\gamma$ (red circles) and light-matter interaction strength $\Omega_{\rm eff}$ (green squares) with varying lateral size of the active exciton volume marked in blue in panel (a). Lines are guides to the eye. (d) Proposed  structure with active two-dimensional MoS$_2$ flake, marked \change{similarly} as in panel (a). (e) Lowest energy polariton mode calculated for the structure shown in (d). (f) An AlGaAs polariton waveguide structure with standing-wave polariton mode pinned to an active exciton region.}
    \label{fig:confinement}
\end{figure*}

One of the outstanding challenges in exciton-polariton physics is the achievement of polariton blockade, a regime where strong interactions result in highly non-classical polariton statistics. This regime occurs when polariton-polariton interactions are much stronger than decoherence. Typically, this is the case when the condition $U/\gamma>1$ is fulfilled with interaction coefficient $U$ and polariton decay rate $\gamma$. However, in typical systems one has $U/\gamma\ll1$. Existing methods and proposals to enhance the interaction strength include strong confinement of the electromagnetic field~\cite{verger2006polariton,Munoz-Matutano2019,Delteil2019,ferretti2012single}, inducing electric dipole moment for excitons~\cite{togan2018enhanced,datta2022highly,dipolar_interaction_citation,liran2024electrically}, and exploiting interactions with electrons~\cite{emmanuele2020highly,ravets2018polaron,tan2020interacting}. In any case, strong confinement of light mode volume seems to be required, but it is difficult to squeeze light much below the wavelength of light. This may be achieved using plasmonic excitations, but at the cost of high loss rates. 

Here we consider using exciton confinement rather than photon confinement to increase the strength of effective interactions between polaritons. Contrary to the common belief, we show that tight photon confinement is not crucial for achieving the polariton blockade regime. It is the exciton confinement that is the determining factor, along with the exciton-exciton interaction coefficient and the decay rate.

In contrast to quantum dots, we consider weak exciton confinement, which we understand as the confinement on \change{a} spatial scales much larger than the exciton size, such that excitons can be still well described as interacting harmonic oscillators, or bosons in the quantum case. Using our theoretical method, we analyze in detail the possibility of obtaining a high ratio of $U/\gamma$ by engineering the structure geometry in a vertical microcavity structure depicted in \textbf{Figure~\ref{fig:confinement}}(a). The structure (not shown in the Figure in its full extent) includes a pair of dielectric Bragg mirrors, each composed of 35 $\rm Al_{0.95}Ga_{0.05}As/Al_{0.2}Ga_{0.8}As$ layer pairs.
Between the mirrors, a $3\lambda/2$ $\rm Al_{0.95}Ga_{0.05}As$ cavity is placed, with a 10\,nm InGaAs QW layer at the antinode of the photonic cavity mode. The transverse size of the QW volume is 0.5$\mu$m and the extent in the direction perpendicular to the Figure plane is 1$\mu$m. Refractive indices of these materials (including attenuation) are available in public databases, so it is possible to model this structure without fitting parameters. This setting is similar as in~\cite{Munoz-Matutano2019}, but we assume no lateral structural confinement for photons. Instead, there is a small volume of "active" material in which the energy of the excitonic excitations is tuned to the energy of the photonic mode, while in the rest of the \change{structure} the exciton energy is out of resonance. Such a weak spatial confinement in quantum wells may be achieved e.g.~by selective interdiffusion method~\cite{zaitsev2007excitons} or additional electrical gates~\cite{thureja2024electrically}.

We determine polariton eigenmodes by solving Maxwell equations  coupled to the material response~\eqref{eq:MaxwellD} including a complex, space dependent dielectric constant. The calculations of eigenmodes are performed with the plane wave admittance transfer method~\cite{dems2005planewave} implemented in the PLaSK software~\cite{Plask} including absorbing boundary conditions with perfectly matching layers. Light-matter coupling strength of the active material was obtained by modeling the system of~\cite{Munoz-Matutano2019} and reproducing the polariton energy splitting. The interaction and exciton decay rates were also taken from~\cite{Munoz-Matutano2019}. The solutions of Maxwell equations allow us to estimate the low-intensity mode energy, decay rate, light-matter interaction and polariton-polariton interaction constant using Equation~\eqref{eq:Hint}. 
To keep computational time in reasonable limits, we determined two-dimensional solutions of Maxwell equations assuming a fixed dimension of $1\mu$m in the direction perpendicular to the Figure plane. While this assumption may not be very accurate taking into account the wavelength of light in dielectric $\lambda/n\approx0.3\mu$m, the physical principle that we aim to demonstrate should hold also in the fully three-dimensional case.

Figure~\ref{fig:confinement}(b) shows the resulting photonic intensity distribution in the lowest energy polariton mode. The photonic part is confined, but on spatial scale much larger than the size of the active QW volume, notice the difference in the scales in Figure~\ref{fig:confinement}(a) and (b). On the other hand, the exciton part is fully confined in the active volume, leading to strong polariton-polariton interactions. According to the results of our theory, a single-mode system can be modeled with  quantum master equation of the form
\begin{align}
\dot{\hat{\rho}} =& \bigg(i \Re (\omega)\left[\hat{\rho}, \hat{a}^\dagger \hat{a}\right] - \Im(\omega) \left(2 \hat{a} \hat{\rho} \hat{a}^\dagger -\left\{ \hat{a}^\dagger \hat{a},\hat{\rho}\right\}\right) \bigg) + \nonumber \\&+ U \left(\hat{a}^\dagger\hat{a}^\dagger\hat{a}\hat{a} + 2 \hat{a}^\dagger\hat{a} + \frac{1}{2}\right),
\end{align}
where $\omega$ is the complex frequency of the polariton mode resulting from Maxwell equations and $U=g \int_V |{\bf X}(\er)|^4 d\er$. However, in this particular structure, the excitons have a two-dimensional character, so we use a corresponding two-dimensional interaction integral $U=g_{\rm 2D} \int_S |{\bf X}_\parallel(\er)|^4 d\er_\parallel$ with ${\bf X}_\parallel(\er)$ being the two-dimensional polarization density.
Indeed, as shown in Figure~\ref{fig:confinement}(c), the reduction of the active layer lateral size leads to the strong increase of the polariton interaction strength $U$. At the same time, the mismatch between photonic and excitonic distributions results in the reduction of the light-matter interaction strength $\Omega$. The loss rate is also slightly increased for smaller layer sizes. However, in a certain range of active layer sizes, the strong interaction regime $U/\gamma>1$ holds together with the strong coupling condition $\Omega/\gamma>1$. 

In Figure~\ref{fig:confinement}(d) we present another proposed structure composed of an atomic layer flake of a two-dimensional material $\rm MoS_2$ sandwiched between $\rm SiO_2/Si_3N_4$ dielectric Bragg mirrors with 25 layer pairs each. Two-dimensional materials are characterized by a very strong light-matter interaction and the silicon-based Bragg mirrors have very low absorption rates, which may be used to fabricate high-Q cavities~\cite{datta2022highly, dufferwiel2015exciton,Kang2023,gu2021enhanced,barachati2018interacting}. The calculated light intensity is shown in Figure~\ref{fig:confinement}(e). In this case, for the $1\mu$m$\times1\mu$m flake size and non-radiative losses rate $\gamma_{\rm NR}=10\mu eV$ we estimate the interaction coefficient to be $U=2.59\mu eV$ and the ratio $U/\gamma=12.95$. 

Finally, Figure~\ref{fig:confinement}(f) shows a possible waveguide polariton structure with a standing-wave polariton mode in the longitudinal direction, based on $\rm AlGaAs$ incorporating a 20\,nm thick GaAs QW volume. The localized active volume acts as a defect, pinning the polariton mode. Due to the total internal reflection, the total decay rate of polaritons in this configuration is only $\gamma\approx 7 \mu$eV assuming nonradiative exciton decay rate of $10 \mu$eV. The interaction coefficient depends on the particular configuration used. Particularly interesting are systems with electric field induced dipolar excitons, where the interaction strength may be of the order of 1~meV~\unit{\micro\meter}$^2$~\cite{liran2024electrically}. In this case, the QW layer of size 1~\unit{\micro\meter}$^2$ results in an interaction coefficient of the order of 1 meV and $U/\gamma\approx 100$. However, even in the case of non-dipolar excitons, ratios of $U/\gamma\approx1$ could be achieved.

\subsection{Engineering nonlocal interactions}

\begin{figure}
    \centering
        \includegraphics[width=0.7\linewidth]{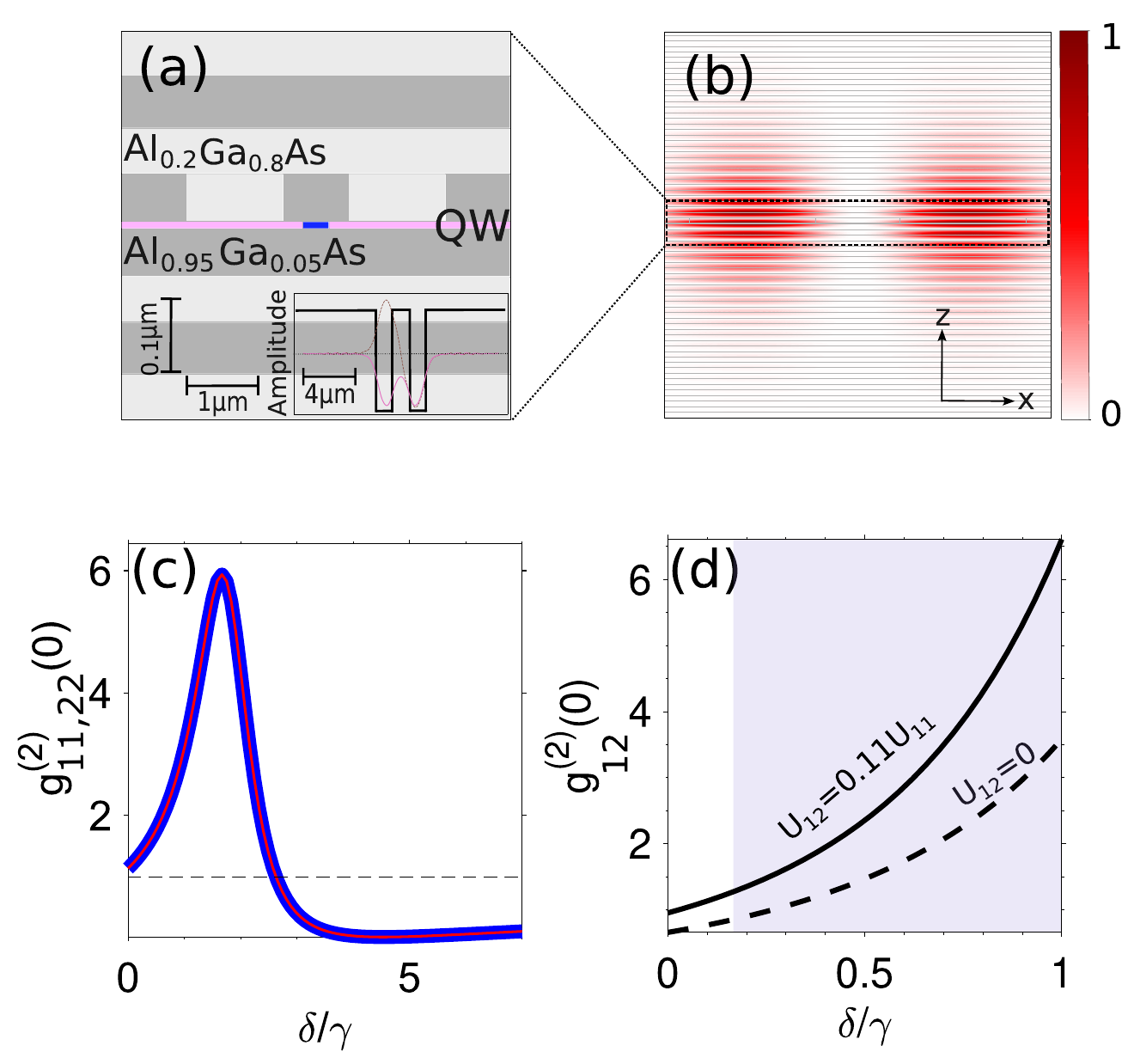}
    \caption{(a) Structure similar to that in Figure~\ref{fig:confinement} but designed for the realization of nonlocal interaction between two polariton modes.  Inset shows amplitude cross-sections of symmetric and antisymmetric modes together with the schematic double well potential resulting from refractive index distribution. (b) Light intensity in the lowest symmetric polariton mode. (c) Second-order equal-time correlation functions $g_{11}^{(2)}(0)=g_{22}^{(2)}(0)$ as a function of detuning in the case of resonant pumping with detuning $\delta$. (d) Cross-site correlation function $g_{12}^{(2)}(0)$ as a function of of detuning (solid line) compared to the case when cross-mode interaction $U_{12}$ is artificially set to zero (dashed line), demonstrating the effect of nonlocal interactions. Shaded area shows the region of the violation of the Cauchy-Schwartz inequality. Calculated parameters of the master equation are $\gamma=9.5 \mu$eV, $J = 47 \,\mu$eV, $U_{11} =U_{22} = 171 \,\mu$eV, and $U_{12} = 0.11 \,U_{11}$.}
    \label{fig:fig3}
\end{figure}

We extend the structure of the previous section to incorporate two confined, but interacting spatial modes, see \textbf{Figure~\ref{fig:fig3}}(a). To this end, we incorporated in the cavity two rectangular regions, filled with $\rm Al_{0.2}Ga_{0.8}As$ material. The refractive index of this material is higher than that of $\rm Al_{0.95}Ga_{0.05}As$, which results in an effective double well potential for photons in the horizontal direction, as depicted schematically in the inset of Figure~\ref{fig:fig3}(a). The position and composition of these regions is not crucial, and in practice the potential could be implemented in other ways, e.g.~using cavity mesas. The inset also depicts the spatial cross-sections of light amplitude in the symmetric and antisymmetric lower polariton modes obtained numerically. The full light intensity distribution of the symmetric mode is shown in Figure~\ref{fig:fig3}(b). To enhance non-local interactions, the active material was positioned between the two potential wells. The horizontal width of the active volume marked in blue is equal to 400~nm.

Symmetric mode (S) and antisymmetric mode (AS) form a lower polariton doublet, shifted from other modes (in particular the upper polariton doublet) by 535~$\mu$eV, which is much larger than the 142 $\mu$eV mode splitting in the doublet. We perform a unitary transformation to the basis of polariton modes localized in the left an right potential well via $\hat{P}_{\rm L,R}=(1/\sqrt{2})(\hat{P}_{\rm S}\mp \hat{P}_{\rm AS})$. After this transformation, master equation takes the form
\begin{align}
\dot{\hat{\rho}} =\frac{1}{i\hbar} [\hat{H},\hat{\rho}] +\sum_{j=L,R} \gamma_j \left(\hat{a}_j \hat{\rho} \hat{a}_j^\dagger -\frac{1}{2}\left\{ \hat{a}_j^\dagger \hat{a}_j,\hat{\rho}\right\}\right),
\end{align}
where
\begin{align}\label{eq:twomode}
    \hat{H} &=  \hbar \omega_{LR} \left(\hat{a}_{\rm L}^\dagger \hat{a}_{\rm L} + \hat{a}_{\rm R}^\dagger \hat{a}_{\rm R}\right) + \\ &+ J \left(\hat{a}_{\rm L}^\dagger \hat{a}_{\rm R} + \hat{a}_{\rm R}^\dagger \hat{a}_{\rm L}\right)+ \frac{1}{2} \Big(U_{LL} \hat{a}_{\rm L}^\dagger \hat{a}_{\rm L}^\dagger \hat{a}_{\rm L} \hat{a}_{\rm L} + \\ & + U_{RR} \hat{a}_{\rm R}^\dagger \hat{a}_{\rm R}^\dagger \hat{a}_{\rm R} \hat{a}_{\rm R} + 2 U_{LR} \hat{a}_{\rm L}^\dagger \hat{a}_{\rm R}^\dagger \hat{a}_{\rm R} \hat{a}_{\rm L}\Big).\nonumber
\end{align}
Here, localized mode energy is $\omega_{\rm LR}=\Re (\omega_{\rm S}+\omega_{\rm AS})/2$, coupling coefficient is $J=\Re (\omega_{\rm S}-\omega_{\rm AS})/2$ and loss coeffcients are $\gamma_{\rm L,R}=-\Im (\omega_{\rm S}+\omega_{\rm AS})$ while the interaction coefficients are $U_{ij}=g \int_V |{\bf X_i}(\er)|^2 |{\bf X_j}(\er)|^2 d\er$ with $U_{LL}=U_{RR}=:U_{11}$ due to mirror symmetry and $U_{LR}=:U_{12}$. We neglected interactions with quantum fluctuations leading to constant frequency shifts. Due to the weakness of dissipative effects compared to mode energies, in this case we can also neglect higher order effects discussed in Sec.~\ref{sec:Lindbladian}.

We calculate the second-order equal-time polariton correlation functions in a steady state under coherent pumping by adding the term $\sum_{j=L,R}\Omega_p(e^{i\omega_p t}\hat{P}_j+e^{-i\omega_p t}\hat{P}_j^\dagger)$ to the Hamiltonian. Since polariton energy is much higher than the light-matter interaction, we work under RWA and transform the Hamiltonian to the frame rotating with pump frequency $\omega_p$, introducing the detuning $\delta=\hbar(\omega_{\rm LR}-\omega_p)$. Steady state intra-mode and inter-mode correlations are shown in Figure~\ref{fig:fig3}(c) and Figure~\ref{fig:fig3}(d), respectively. The system exhibits standard bunching/antibunching behavior when crossing the resonance~\cite{drummond1980quantum}. Interestingly, cross-mode correlations between left and right polariton modes $g_{12}^{(2)}(0)$ exhibit also a strong dependence on detuning, showing strong preference for simultaneous population in both modes (positive cross-correlation) at large detunings. The range of parameters for which the Cauchy-Schwartz inequality is violated, marking the emergence of \change{nonclassical correlations}, is marked with shaded area. This phenomenon is significantly boosted by non-local interactions, as evident from the comparison to the results of simulations where the cross-interaction term was artificially removed, see dashed line in Figure~\ref{fig:fig3}(d).  This illustrates that engineering nonlocal interactions can have a substantial effect on polariton statistics in nanostructures, and could be used to construct sources of quantum light.

\section{Conclusions}

In conclusion, we introduced a method for obtaining a quantum model corresponding to a confined coupled light-matter system, both in the conservative and dissipative regime, with no assumptions on the system geometry. The method is based on quantization in normal modes basis, and allows to treat without approximations systems that are not well described by the  Hopfield model. It requires solving the classical limit equations to determine the complete form of the quantum model in an optimal basis, with no fitting parameters. Even though the method is based on diagonalization of a quadratic system, we showed how it can be used in practice to describe non-quadratic, interacting systems. We proposed practical applications resulting in the increase of effective polariton interactions and engineering non-local interactions in semiconductor exciton-polariton structures. 

In comparison to previous approaches~\cite{huttner1992quantization,drummond1990electromagnetic,Santos,knoll1992quantum,milonni1995field,PhysRevA.53.1818,PhysRevA.57.4818,PhysRevA.57.3931,bechler1999quantum,PhysRevA.95.023831,raymer2020quantum,suttorp2004field,bhat2006hamiltonian}, the main advantage of our treatment is the simplicity of calculations and of the resulting model. We also emphasize that the applicability of the presented method is much broader than the context presented here. For example, one can go beyond the assumption of small emitter size, and the assumption of infinite emitter mass, as long as the classical description of the system is known. It is also possible to add to the model interactions with other relevant excitations, such as phonons. As long as the system is well described as a collection of bosonic modes, the quantum-to classical correspondence can be used to determine the form of the quantum model. Moreover, even when one needs to describe modes which are not bosonsic (for example, spin systems), analogous transformations may be found~\cite{ProsenSpin}.

The presented theoretical method can be used to engineer light-matter systems that include semiconductor nanostructures and cold atom systems. Possible practical applications include engineering systems for quantum simulations~\cite{boulier2020microcavity}, where reproducing a given, precisely defined quantum model is of great importance, and designing sources of quantum light. Finally, our method can provide better understanding of nonconventional polariton systems that are not well described within simplified theoretical models such as the Hopfield model.

% Text: Please use section headings and subheadings as specified below. For communications, all section headings apart from Experimental Section should be removed
% Please make the first reference to a display item bold: \textbf{Figure 1}
% Do not abbreviate Figure, Equation, etc.; display items are always singular, i.e., Figure 1 and 2.
% Equations are always singular, i.e., Equation 1 and 2, and should be inserted using the {equation} environment, not as graphics
% Please do not use footnotes in the text, additional information can be added to the Reference list.

% \section{First Section}

% \subsection{First Subsection}

% \subsubsection{First Sub Subsection}

% \threesubsection{First lowest-level subsection}

% \section{Conclusion}

% Experimental section

% \section{Experimental Section}
% \threesubsection{First part of experimental section}\\
% \threesubsection{Second part of experimental section}\\

% \medskip
% \textbf{Supporting Information} \par %Please delete the Suppporting Information statement if it is not applicable. Please supply Supporting Information in another file. Supporting information should not be provided in .tex format
% Supporting Information is available from the Wiley Online Library or from the author.

%Acknowledgements
\medskip
\textbf{Acknowledgements} \par %delete if not applicable))
This project received funding from the European Union’s Horizon Europe research and innovation programme under grant agreements No.~ID 101115575 (Q-ONE) and ID 101130384 (QUONDENSATE). AO acknowledges support from the National Science Center, Poland, project No. 2024/52/C/ST3/00324.

\medskip
\textbf{Conflict of Interest} \par
The authors declare no conflict of interest.

\medskip
\textbf{Data Availability Statement} \par
The data and source codes that support the findings of this study \change{are available at \cite{rahmani_zenodo}}.

% References
\medskip

% Use the following code if you wish to generate your bibliography with BibTeX;
% replace the string "MSP-template" below with the name(s) of
% the BibTeX data base(s) you want to use.
% The resulting bibliography-output (the content of the .bbl file)
% must be pasted back into this file before submission.
% Please also include your BibTeX data base file(s) in your submission
% so that we can re-run BibTeX if necessary.
%
%\bibliographystyle{MSP}
%\bibliography{mybib}

\end{document}